\documentclass[twocolumn,showpacs,aps,nofootinbib]{revtex4}
\usepackage{color}   
\usepackage{graphicx}
\usepackage{dcolumn}
\usepackage{bm}
\usepackage{slashed}

\begin{document}

\def\bb    #1{\hbox{\boldmath${#1}$}}

\title{Bose-Einstein Interference in the Passage of a Jet in a Dense
  Medium$^\dag$ \footnote[0]{ $^\dag$Based in part on a talk presented
    at the Fourteenth Meeting of the Chinese Nuclear Physical Society
    in Intermediate and High Energy Nuclear Physics, Qufu, Shandong,
    China, October 20-24, 2011.} }

\author{Cheuk-Yin Wong}

\affiliation{Physics Division, Oak Ridge National Laboratory, 
Oak Ridge, TN 37831}

\date{\today}

\begin{abstract}

When a jet collides coherently with many parton scatterers at very
high energies, the Bose-Einstein symmetry with respect to the
interchange of the virtual bosons leads to a destructive interference
of the Feynman amplitudes in most regions of the momentum transfer
phase space but a constructive interference in some other regions of
the momentum transfer phase space.  As a consequence, the
recoiling scatterers have a tendency to come out collectively along
the incident jet direction, each carrying a substantial fraction of
the incident jet longitudinal momentum.  The manifestation of the
Bose-Einstein interference as collective recoils of the scatterers
along the jet direction may have been observed in  angular
correlations of hadrons associated with a high-$p_T$ trigger in
high-energy heavy-nuclei collisions at RHIC and LHC.

\end{abstract}

                                                                         
\maketitle

\section {Introduction}

Recently at RHIC and LHC, angular correlations of produced hadron
pairs in AuAu, PbPb, and $pp$ collisions have been measured to obtain the
yield of produced pairs as a function of $\Delta \phi$ and $\Delta
\eta$, where $\Delta \phi$ and $\Delta \eta$ are the azimuthal angle
and pseudorapidity differences of the produced pair.  The correlations
appear in the form of a ``ridge'' that is narrow in $\Delta \phi$ at
$\Delta \phi\sim 0$ and $\Delta \phi\sim \pi$, but relatively flat in
$\Delta \eta$.  They have been observed in high-energy AuAu collisions
at RHIC by the STAR Collaboration
\cite{Ada05,Ada06,Put07,Bie07,Wan07,Bie07a,Abe07,Mol07,Lon07,Nat08,Fen08,Net08,Bar08,Dau08,Ray08a,Ket09,STA11,Lee09,Tra08},
the PHENIX Collaboration \cite{Ada08,Mcc08,Che08,Jia08qm,Tan09}, and
the PHOBOS Collaboration \cite{Wen08}, with or without a high-$p_T$
trigger \cite{Ada06,Tra08,Dau08,Ray08a,Ket09,STA11}.  They have also been observed in
$pp$ and PbPb collisions at the LHC by the CMS Collaboration 
\cite{CMS10,CMS11,CMS12}, the ATLAS
Collaboration \cite{ATL12}, and the ALICE Collaboration \cite{ALI12}.

Subsequent to the observation of the $\Delta \phi$-$\Delta \eta$ angular correlations, 
a momentum kick model \cite{Won07,Won08ch,Won08,Won08a,Won09,Won09a,Won09b,Won11,Won12}
was put forth to explain the ``ridge''
phenomenon, along with many other models \cite{Shu07,Vol05,Chi08,Hwa03,Chi05,Hwa07,Pan07,Dum08,
Gav08,Gav08a,Arm04,Rom07,Maj07,Dum07,Miz08,Jia08,
Jia08a,Ham10,Dum10,Wer11,Hwa11,Chi10,Tra10,Tra11a,
Tra11b,Tra11c,Sch11,Pet11,Arb11,Aza11,Hov11,Bau11,Che10,Dre10,Lev11}.
 The
model identifies ridge particles as medium partons, because of the
centrality dependence of the ridge particle yield and the similarity between the temperature and baryon/meson ratio of ridge particles with
those of the bulk medium.  The model assumes that these
medium partons have an initial rapidity plateau distribution, and when
they suffer a collision with a jet produced in the collision,
 they
receive a momentum kick along the jet direction, which will be designated as the longitudinal direction.
 The
physical contents of the momentum kick model lead to features
consistent with experimental observations: (i) the longitudinal 
momentum kicks from the jet to the medium partons 
 along the jet direction give rise to the $\Delta \phi \sim 0$ correlations on
the near-side
\cite{Ada05,Put07,Won07,Won08,Won08ch,Won08a,Won09}, (ii) the
medium parton initial rapidity plateau distribution  
shows up as a
ridge along the $\Delta \eta$ direction
\cite{Ada05,Put07,Won07,Won08,Won08ch,Won08a,Won09,Won09b,Won11},
(iii) the displaced parton momentum distribution of the kicked medium
partons leads to a peak at $p_T\sim 1$ GeV/c  in the $p_T$ distribution of the kicked
partons \cite{Tra08,Won08,Won11}, and (iv) the kicked medium partons
therefore possess a correlation of $|p_{1T}| \sim |p_{2T}|$$\sim$1
GeV, without a high-$p_T$ trigger, for both the near side and the away
side \cite{Tra08,Won11}.  The model has been successful in describing
extensive sets of triggered associated particle data of the STAR
Collaboration, the PHENIX Collaboration, and the PHOBOS Collaboration,
over large regions of $p_t$, $\Delta \phi$, and $\Delta \eta$, in many
different phase space cuts and $p_T$ combinations, including
dependencies on centralities, dependencies on nucleus sizes, and
dependencies on collision energies
\cite{Won07,Won08ch,Won08,Won08a,Won09,Won09a,Won09b,Won11}.
  
The phenomenological success of the momentum kick model raises
relevant questions about the theoretical foundations for its basic
assumptions.  While the rapidity plateau structure of the medium parton
distribution may have its origin in the Wigner function of particles
produced in the fragmentation of a string (or flux tube)
\cite{Won09a}, or in a color-glass condensate \cite{Dum08}, the origin
of the postulated longitudinal momentum kick along the jet direction in the model poses an
interesting puzzle.  If each of the jet-(medium parton) collision were
a two-body elastic collision, it would lead only to a
dominantly forward scattering and a small longitudinal momentum
transfer to the medium scatterers along the jet direction.  The hard longitudinal momentum kick (of order GeV) 
received by the medium scatterers
along the jet direction in the
momentum kick model is unlikely to originate from incoherent two-body
collisions.  

From an observational viewpoint, the experimental observation that
many medium partons come out along the jet longitudinal direction with $\Delta
\phi\sim 0$ implies the presence of collective recoils of the medium
partons, each of which must have acquired a substantial longitudinal
momentum transfer (momentum kick) along the jet direction from the
jet.  What is the origin of such collective recoils, in the
interaction of the jet with medium partons?

As the high-$p_T$ trigger particles under experimental consideration have a momentum of order a few GeV
\cite{Ada05,Ada06,Put07,Ada08,Mcc08,Che08,Jia08qm,CMS11,CMS12,ATL12,ALI12},  we therefore consider incident jets 
with an initial momentum of order 10 GeV.
Strictly speaking, jets with an initial momentum of order
10 GeV should be more 
appropriately called ``mini-jets" \cite{Wan92}.   For brevity of nomenclature, we continue to use the term ``jet" to represent mini-jets.   

 In the collision of a jet $p$=$(p_0,{\bb p})$
with $n$ medium partons, how dense must the medium be for the
multiple collisions to become a single coherent (1+$n$)-body
collision, instead of a sequence of $n$ incoherent 2-body collisions?
It is instructive to find out the conditions on the medium density and
the energy of the incident jet that determine whether the set of
multiple collisions are coherent or incoherent.  For
such a purpose, we consider a binary collision between the incident
fast particle $p$ and a medium scatterer $a_i$ with the exchange of a
boson, in the medium center-of-momentum frame.  The longitudinal
momentum transfer $q_z$ for the binary collision can be obtained from
the transverse momentum transfer $q_T$ by $q_z \sim q_T^2/2p_0$.  The
longitudinal momentum transfer is associated with a longitudinal
coherence length $\Delta z_{\rm coh}\sim \hbar /q_z \sim 2\hbar
p_0/q_T^2$ that specifies the uncertainties in the longitudinal
locations at which the virtual boson is exchanged between the fast
particle and the scatterer.
 The time it takes
for the fast particle to travel the distance of the longitudinal
coherence length, $ 2p_0 \hbar/q_T^2 c$, can also be called the
virtual boson formation time.

The nature of the multiple scattering process can be inferred by
comparing the longitudinal coherence length $\Delta z_{\rm coh}$ with
the mean free path $\lambda$ of the jet in the dense medium that
depends not only on the density of the medium but also on the binary
collision cross section.  

If $\Delta z_{\rm coh} \ll \lambda$, then a
single binary collision is well completed before another binary
collision begins, and the multiple collision process consists of a
sequence of $n$ incoherent two-body collisions.
On the other extreme, if 
\begin{eqnarray}
\Delta z_{\rm coh} \sim 2\hbar p_0/q_T^2
\gg \lambda,
\label{1}
\end{eqnarray}
then a single binary collision is
not completed before another one begins, and the multiple collision
process consists of a set of coherent collisions as a single
(1+$n$)-body collision.  For a set of initial and final states in such
a coherent (1+$n$)-body collision, there are $n!$ different
trajectories in the sequence of collisions along $\Delta z_{\rm coh}$
at which various virtual bosons are exchanged.  By Bose-Einstein
symmetry, the total Feynman amplitude is then the sum of the $n!$
amplitudes for all possible interchanges of the exchanged virtual
bosons.  The summation of these $n!$ Feynman amplitudes and the
accompanying interference constitute the Bose-Einstein interference in
the passage of the fast particle in the dense medium.

In high
energy central collisions between heavy nuclei such as those at RHIC
and LHC, both jets and dense medium are produced after
each collision.  The jets will collide with partons in the dense
medium, and these collisions may satisfy the condition for coherent collisions.
In a binary collision at RHIC and LHC, the longitudinal coherent length $\Delta z_{\rm coh}$ is
of order 25 fm, for a typical transverse momentum transfer of
$q_T$$\sim$0.4 GeV/c  from a jet of momentum $p_0$$\sim$10 GeV/c  to a
medium parton.  The
longitudinal coherent length $\Delta z_{\rm coh}$ is much greater than
the radius $R$ of a large nucleus.  On the other hand, the away side
jet is quenched by the dense medium in the most central AuAu
collisions at RHIC and LHC
 \cite{Bjo82,Gyu94,Bai96,Wie00,Gyu01,Djo04,Adl04}, and the near-side
jet collides with about 4-6 medium partons \cite{Won07}-\cite{Won11}.
Therefore, one can infer that the mean-free path $\lambda$ for the
collision of the jet with medium partons is much smaller than the
nuclear radius $R$.    In
high-energy central nuclear collisions at RHIC and LHC, the collision of a fast jet with medium scatterers satisfy the following condition 
\begin{eqnarray}
\Delta z_{\rm coh} \sim 2\hbar p_0/q_T^2 \gg R
\gg \lambda.
\label{2}
\end{eqnarray}
As a consequence, the multiple collision process  constitutes a set of coherent collisions.
There will be Bose-Einstein interference effects in the passage of the jet in the dense
medium.

In this connection of Bose-Einstein interference, we note that
Bose-Einstein interference effects in high-energy QED and
QCD collisions have been observed previously by many workers
\cite{Che69,Che87,Fen96,Fen97,Lam96a,Lam97,Lam97a,Lam97b}.  In the
emission or absorption of $n$ identical bosons from an energetic
fermion in Abelian and non-Abelian gauge theories leading to an
on-mass-shell final fermion, the Bose-Einstein symmetry with respect
to the interchange of virtual bosons leads to the sum of a the set of $n!$
Feynman amplitudes that turns out simply to be a product of delta
functions.  These distributions may be thought of as peaked
interference patterns produced by the coherent addition of various
symmetrized broad Feynman amplitudes \cite{Lam97b}.  In the
collision of two fermions, the sum of the ladder and cross-ladder
diagrams also exhibits remarkable Bose-Einstein interference leading
to similar products of delta functions and the eikonal approximation
\cite{Che69,Che87,Fen96,Fen97,Lam96a,Lam97,Lam97a,Lam97b}.

We would like to study similar interference effects for the case of
the multiple collisions of a jet parton with many parton scatterers.
We wish to explore whether such a Bose-Einstein 
interference may be the origin of the longitudinal momentum kick.  As
the interference arises from the coherent collision of more than one
particle, it is clearly a many-body effect that 
occurs with two or more scatterers.

The interference of Feynman amplitudes is only one of the effects of
coherent collisions.  There is another important effect that accompanies the coherent
(1+$n$)-body collision and changes the nature of the collision
process.  It shows up as an increase in the number of degrees of
freedom. In a sequence of $n$ incoherent two-body collisions or quasi-elastic collisions, there
are two degrees of freedom in each two-body collision, which can be
chosen to be ${\bb q}_{iT}=(q_{iT},\phi_{i})$.  The sequence of $n$
incoherent or quasi-elastic two-body collisions contains only $2n$ transverse momentum
transfer degrees of freedom, $\{ {\bb q}_{1T},{\bb q}_{2T},{\bb
  q}_{3T},...,{\bb q}_{nT}\}$.  The longitudinal momentum transfer $q_{iz}$ in
each individual jet-parton collision is a dependent variable,
depending on the corresponding transverse momentum transfer as 
$q_{iz} \sim |\bb q_{iT}|^2/2p_0$.
In contrast, in the case of coherent
collisions of the jet with $n$ particles, there are $3n-1$ degrees of
freedom, after the $3(n+1)$ degrees of freedom are reduced by the
constraints of the conservation of energy and momentum.  The degrees
of freedom increases from $2n$ for incoherent collision and quasi-elastic collisions to $3n-1$ for
coherent collisions.  We can choose the $3n -1$ independent variables
to be $\{ {\bb q}_{1T}q _{1z},{\bb q}_{2T}q _{2z},{\bb q}_{3T}q
_{3z},...,{\bb q}_{nT}q _{nz}\}$ for coherent collisions, subject to a
single condition of an overall energy conservation.  Thus in the case
of coherent collisions, the set of longitudinal momentum transfers,
$\{ q_{iz}, i=1,2,...,n\}$, can also be independent variables with
their own probability distribution functions, depending on the sum of $n!$ symmetrized
Feynman amplitudes.  The recoiling scatterers share the longitudinal
momentum of the incident jet.  The $q_{iz}$ degrees of freedom allow
the medium partons to acquire substantial fractions of the
longitudinal momentum of the incident jet, as we shall demonstrate in
Section V of this paper.

In the problem of the passage of a jet in a dense medium, the multiple collision process  of the jet with medium partons is usually examined in 
the potential model  \cite{Gyu94,Bai96,Wie00,Gyu01,Djo04,Ova11,Meh11,Arm12}, as  in the Glauber theory \cite{Gla59}.
It should be noted that the occurrence of coherent collision discussed here coincide with the condition for multiple scattering in the Glauber theory given by Eq.\ (173) of \cite{Gla59} as
\begin{eqnarray}
 p_0 a^2/\hbar \gg R,
\label{gla}
\end{eqnarray}
where $R$ is the dimension of the medium, and $a$ is the range of the interaction which can be related to the average transverse momentum transfer by $a\sim \hbar/q_T$.
Thus the Galuber's condition for multiple scattering is
\begin{eqnarray}
\hbar {p_z}/{q_T^2} \gg R.
\end{eqnarray}
As the dimension of the medium is presumably much greater than the mean free path $\lambda$ between collisions, condition (\ref{gla}) for Glauber's multiple scattering is consistent with 
\begin{eqnarray}
\hbar {p_z}/{q_T^2} \gg R \gg \lambda
\label{5a}
\end{eqnarray}
that is nearly the same as 
the coherent condition (\ref{2}) we discussed here, except for an unimportant difference of a factor of two.

The near-coincidence of the conditions for coherent collisions of
Eqs.\ (\ref{2}) and (\ref{5a})  indicates that the totality of coherent collisions can be further divided  according to the degree of incident jet  longitudinal momentum losses. 
There are elastic and quasi-elastic coherent collisions with nearly no scatterer recoils and little loss of the incident jet longitudinal momentum
that are at the center of Glauber-type potential model theories.  There are also coherent collisions in which scatterers can undergo longitudinal recoils and the incident jet can lose a substantial longitudinal momentum arising from the  recoils of scatterers.
The elastic and quasi-elastic coherent collision processes can be adequately studied 
in the potential model \cite{Gyu94,Bai96,Wie00,Gyu01,Djo04,Ova11,Meh11,Arm12,Gla59}, 
with the assumption that the scatterers suffer approximately no recoils and appear in the form of potential centers around which the projectile traverses.  

We wish to investigate coherent collisions in the entire domain of longitudinal scatterers recoils and the corresponding 
incident jet  longitudinal momentum loss due to these recoils.
For a general treatment of longitudinal  recoils
of scatterers,
 the Glauber-type potential model cannot really be used. 
It is inadequate
 because it is restricted to elastic and quasi-elastic collisions and 
does
not allow the scatterer longitudinal recoils to be independent degrees
of freedom. 
For well-founded reasons which we shall discuss in detail in Section II, we shall forgo the potential model but turn to the Feynman amplitude approach to study the 
longitudinal 
recoils of the scatterers in  a general treatment of coherent collisions.

In the remaining parts of the manuscript, we shall consider implicitly
only coherent multiple collisions of the jet with medium partons
unless indicated otherwise.    In Section III, we
examine (coherent) multiple collisions of an energetic fermion with
fermion scatterers in the Abelian gauge theory to discuss how the
Bose-Einstein symmetry in the Feynman amplitudes can lead to a
destructive interference over most regions of phase space but a
constructive interference in some other regions of phase space.  In
Section IV, we study the cross section for such a multiple collision
process.  In Section V, we examine the consequences of the constraints on the
recoils of the scatterers and find that the scatterers tend to come
out collectively along the incident jet direction, each scatterer
acquiring approximately $1/2n$  fraction of the longitudinal momentum of the
incident jet.  In Section VI, we examine the case for the coherent
collision of an energetic fermion with fermion scatterers in the
non-Abelian gauge theory.  In Section VII, we generalize our
considerations to the collision of a gluon jet on quark scatterers.
In Section VIII, we examine the collision of a gluon or a fermion jet
on gluon scatterers.  In Section IX, we study the longitudinal
momentum distribution for gluon scatterers and distinguish it from the
longitudinal momentum distribution for quark scatterers.  In Section
X, we examine the signatures of the Bose-Einstein interference in the
passage of a jet in a medium and compare these signatures with
experimental data.  In Section XI, we present our conclusions and
discussions.

\section{
The Potential Model
versus  the Feynman Amplitude Approach}

The potential model of multiple collisions
\cite{Gyu94,Bai96,Wie00,Gyu01,Djo04,Ova11,Meh11,Arm12,Gla59}
 has been
a useful tool to study the energy loss of a jet as it  passes through a
dense medium in elastic and quasi-elastic scatterings.  It can also be adequately used to study the longitudinal momentum transfers $q_{iz}$  from the jet to the scatterers in these elastic and quasi-elastic  processes.  The potential model gives the longitudinal recoil momentum 
\begin{eqnarray}
q_{iz}\sim |\bb q_{iT}|^2/2p_0,
\label{7}
\end{eqnarray}
for each jet-scatterer collision, in either coherent or incoherent collisions.

We shall see that  in the Feynman amplitude approach the probability  distribution for the longitudinal recoil  $q_{iz}$ includes not only the probability for quasi-elastic processes with little  scatterers  longitudinal recoils but also
a significant probability for  the scatterers to recoil with a substantial fraction of the incident jet longitudinal momentum.
It is therefore important to inquire which of the two different approaches should be used to
investigate  the longitudinal recoils of the scatterers
in a general treatment  of coherent collisions that are at the focus of our attention.

It is instructive to review how the longitudinal recoils of the scatterers are determined in the two different  approaches.
In the potential model \cite{Gyu94,Bai96,Wie00,Gyu01,Djo04,Ova11,Meh11,Arm12,Gla59}, the medium parton
scatterers are assumed to be fixed in space and the scatterings between the incident particle and the medium scatterers are represented by potentials with the positions
of the scatterers as static centers, as in the Glauber model for
potential scattering \cite{Gla59}.  By using {\it static} scatterer
centers, the range of applicability of the potential model is limited
to elastic or quasi-elastic scatterings with small momentum transfer $\bb q_i$.  Furthermore, we can trace how
$\bb q_i$=$(\bb q_{iT},q_{iz})$ is determined in the potential model, in the collision of the
jet with the $i$th scatterer.  After passing through the potential
generated by the $i$th scatterer, the trajectory of the incident jet
is obtained and the transverse momentum transfer $\bb q_{iT}$
determined.  The longitudinal momentum transfer $q_{iz}$ from the jet
to the scatterer is subsequently calculated to ensure longitudinal
momentum conservation. Hence the longitudinal momentum transfer from
the jet to the $i$th scatterer is determined as $q_{iz} \sim \bb
q_{iT}^2/2 p_0$,  as given in Eq.\ (\ref{7}).  As the longitudinal momentum transfer $q_{iz}$ is
determined completely by $\bb q_{iT}$ and $p_0$, it is no longer an
independent dynamical variable.  It is in effect a frozen and
dependent quantity.  Not only is the potential model limited to
scattering with small momentum transfers $\bb q_i$, the model is also
limited to cases in which the longitudinal momentum transfers $q_{iz}$
from the jet to the scatterers are no longer independent dynamical
variables.

In the potential model with many scatterers, the total potential is
the sum of all scatterer potentials.  In the coherent collisions limit,
the wave function of the incident jet becomes
\cite{Gla59}
\begin{eqnarray} 
\label{psi}
\psi(\bbox{b} z)=\exp\{ik(z-z_0) +i \chi_{\rm tot} (\bbox{b},z) \},
\end{eqnarray}
where the phase $\chi_{\rm tot} (\bbox{b},z)$ is the sum of the
scattering phases from scatterer potentials $V_i$ centered at
$\{\bb b_i, z_i\}$,
\begin{eqnarray} 
\chi_{\rm tot} (\bbox{b},z)=
-\frac{i}{hv} \int_{z_0}^z dz'   \sum_{i=i}^n V_i( {\bb b} -{\bb b_i}, z'-z_i).
\end{eqnarray}
The width of the transverse momentum distribution of the incident jet is
broadened by collisions with scatterers.  From the broadening of the
transverse momentum distribution associated with the transverse momentum
transfer $\bb q_{iT}$ due to the interaction of the jet with the $i$th
scatterer, one infers that the corresponding longitudinal momentum
transfer to the $i$th scatterer is given again by $q_{iz}\sim |\bb
q_{iT}|^2/2p_0$ as in Eq.\ (\ref{7}) in the coherent collision limit.

From the above review, we recognize that in the
potential model, all recoils of the scatterers have been  assumed to be zero 
initially to allow a static description of the centers of the
scatterers.   The scatterer longitudinal recoils are only subsequently corrected
as an appendage to the dynamics of the transverse deflection of the
incident jet.  Such a potential model may be
adequate for 
incoherent collisions and for quasi-elastic coherent collisions
 in which the longitudinal momentum transfers to the scatterers are
very small.   

As the longitudinal recoils of the scatterers are at the focus of our
attention, it is important to realize that the potential model cannot be used to examine the longitudinal recoils of the scatterers in a general treatment where one wants to explore the behavior of the probability distribution  for the  scatterer longitudinal momentum
transfer over a large domain, or in regions where there may be  a high
probability for recoiling scatterers to share substantial fractions of
the longitudinal momentum of the incident jet. 
A general treatment of
 coherent collisions
necessitates the use of
the 
longitudinal recoils of the scatterers
as
independent dynamical variables.   The longitudinal  recoils of the scatterers are 
however
not allowed
to be independent  dynamical variables in the potential model. 
This leads us to forgo the potential model and to turn to the use of
Feynman amplitudes for coherent collisions.

In the Feynman amplitude approach such as given in what follows for a set of the initial momenta of the jet and scatterers,  the  $(n+1)$  final momenta involved in
the collision process are dynamical variables.  The Feynman
amplitudes give the probability amplitudes for various reaction
channels as a function of these dynamical variables.  In particular, the
longitudinal momenta of the recoiling scatterers and the corresponding
longitudinal momentum transfers from the jet to the scatterers are
dynamical variables, in contrast to the potential model in which they
are frozen and dependent quantities. 
Feynman amplitudes with
explicit scatterer momenta on the external legs are the proper tools to examine
the probability amplitudes as a function of the longitudinal recoils of the scatterers.

In the Feynman amplitude approach, the momenta of recoiling scatterers are
intimately linked together with the 
momentum of the incident jet in various Feynman
amplitudes and their corresponding symmetrized $n!$ permutations.  The $n$
scatterers and the jet comprise an $(n+1)$-body system,
with $3(n+1)$ degrees of freedom constrained by the four-dimensional
energy and momentum conservation.  Among the $3(n+1)$ degrees of
freedom are the transverse and longitudinal momentum transfers from the jet to  the scatterers,
$(\bb q_{iT}, q_{iz}),~i=1,..,n$.  As the recoiling scatterers 
and  the recoiling jet are tied together by the coherent
collision, 
they can share the initial  longitudinal
momentum of the jet.  
We shall find 
from  the Feynman 
amplitude approach that $\langle q_{iz} \rangle $
can be a substantial fraction of
the  longitudinal momentum of the incident jet (see Section IV).  The
potential model  in contrast gives the result  $q_{iz}=|\bb
q_{iT}^2|/2p_0$  that 
cannot be relied upon for a general coherent collision,
because (i) the potential model 
confines the system to regions of small $q_{iz}$, (ii) 
as it has been designed for elastic and quasi-elastic processes, the potential model does not allow $q_{iz}$ to be  independent dynamical variables, and (iii) as a consequence,  the potential model precludes the exploration into other $q_{iz}$  regions  where the probability 
distribution as a function of $q_{iz}$ 
 may be large in coherent collisions.

In the general treatment of  coherent collisions with the jet,
we are therefore justified to forgo the potential model and use the Feynman amplitude approach  to study the transverse and longitudinal recoils of  parton scatterers.

\section{Bose-Einstein Interference of the Feynman Amplitudes}

We consider a jet $p$ passing through a dense medium and making
(coherent) multiple collisions with medium partons.  The assembly of
medium partons has an initial momentum distribution.  We choose to
work in the center-of-momentum frame of the  parton
scatterers $\{a_1,a_2,...,a_n\}$.  We select the longitudinal $z$-axis
to be along the momentum of the incident jet.

As an illustration of the salient features of the interference of the
Feynman amplitudes, we consider first multiple collisions of
a fermion $p$ with rest mass $m$ and two fermion scatterers $a_1$ and $a_2$ with rest masses $m_1$ and $m_2$ in
$p+a_1+a_2 \to p'+a_1'+a_2'$.   We study this problem in the
Abelian gauge theory in this section and in the non-Abelian gauge
theory in Section VI. We shall consider the collision of gluons in
Section VII.

\begin{figure} [h]
\includegraphics[scale=0.45]{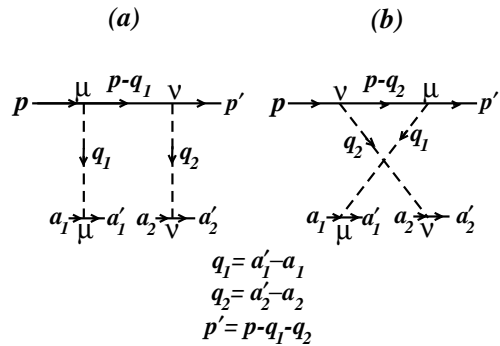}
\caption{ Feynman diagrams for the collision of a fast fermion $p$ with medium
  fermions $a_1$, $a_2$, with the emission and absorption of virtual
  bosons of momenta $q_1$ and $q_2$.  }
\end{figure}
Using the Feynman rules as in Ref.\ \cite{Che87},
the Feynman  amplitude for diagram 1(a) is
given by  \cite{Note,Itz80}
\begin{eqnarray}
M_a &= &- g^4 {\bar u}({\bb p}')
  \gamma_\nu \frac{1}{\slashed p - \slashed q_1-m+ i \epsilon'} 
\gamma_\mu u({\bb p})
\nonumber\\
&\times&
\frac{1}{q_2^2}  {\bar u}({\bb a}_2')
\gamma_\nu u({\bb a}_2)
\frac{1}{q_1^2}  {\bar u}({\bb a_1}')
\gamma_\mu u({\bb a}_1).
\end{eqnarray}

If the spatial separation between the scatterers is so large such that $\lambda \gg \Delta z_{\rm coh}$  the fast particle is nearly on the mass shell 
after the first collision and the
diagram 1(a) suffices.  It  
can be cut into two disjoint pieces.   The collision process consists effectively of a sequence of two two-body collisions.  We shall show in Section IV that for a  ladder diagram of the  
 type  shown in 
Fig.\ 1(a)  for quasi-elastic collision, with intermediate fast particles assumed to be on the mass shell and $q_{i0}\sim 0$, the (1+$n$) body cross section is a product of $n$ two-body cross sections. 

On the other hand, if the collision  is characterized by
$\lambda \ll \Delta z_{\rm coh}$ the
 collision process is 
in a 
coherent
(1+2)-body collision.  
There is an additional Feynman amplitude $M_b$
for diagram 1(b), obtained by making a symmetrized permutation of the
bosons in diagram 1(a),
\begin{eqnarray}
M_b &= &- g^4 {\bar u}({\bb p}')
  \gamma_\mu \frac{1}{\slashed p - \slashed q_2-m+ i \epsilon'} 
\gamma_\nu u({\bb p})
\nonumber\\
&\times& 
\frac{1}{q_2^2}  {\bar u}({\bb a}_2')
\gamma_\nu u({\bb a}_2)
\frac{1}{q_1^2}  {\bar u}({\bb a_1}')
\gamma_\mu u({\bb a}_1).
\end{eqnarray}
The trajectories for diagram 1(a) and 1(b) are both possible paths in
a coherent collision, leading from a set of initial states to a
set of final states.  By Bose-Einstein symmetry, the total amplitude
$M$ for coherent collisions is the symmetrized sum of $M_a$ and $M_b$.

We  consider the high-energy limit and assume the conservation of helicity with 
\begin{eqnarray}
\{ p_0, |{\bb p}| ,p_0', |{\bb p}'|\} \gg \{ |{\bb a_i }|, q_0,|{\bb q_i }|\}  \gg m, {\rm ~for~}
i=1,2.\label{2a} \nonumber
\end{eqnarray}
We shall later find that $\langle q_{iz}\rangle/p_{z}$ is of order 1/2$n$.  The above high-energy limit of 
$\{ p_0, |{\bb p}| ,p_0', |{\bb p}'|\} \gg \{  q_0,|{\bb q_i }|\}$ provides a reasonable approximation to the gross features of the collision processes.  Higher order corrections to this approximation will lead to refinements that will be the subjects of future investigations.   

In this high-energy limit, we have approximately \cite{Che87}
 \begin{eqnarray}
&&
\hspace*{-0.4cm}{\bar u}({\bb a}') \gamma_\mu u({\bb a})
\!\sim\!
\sqrt{\frac{a_0+m}{{a_0'+m} }}
\frac{a_\mu'}{2m}
\!+\! \sqrt{\frac{a_0'+m}{a_0+m} }
\frac{a_\mu}{2m}
 \equiv  
\frac {{\tilde a}_\mu}{m}, \label{5} \\
&&\frac{1}{\slashed p - \slashed q_1-m +i \epsilon'} 
\gamma_\nu u(p)
\sim  - \frac{\slashed p - \slashed q_1+m }{2p \cdot q_1 - i \epsilon} 
\gamma_\nu u(p),
\\
&&{\bar u}(p) \gamma_\nu  (\slashed p - \slashed q_1+m)   \gamma_\mu u(p)
\sim   \frac{ 2 p_{\nu}p_{\mu}}{m}, 
\end{eqnarray}
where $\epsilon$ is a small positive quantity.  We shall be interested
in the case in which the fermion $p'$ after the collision is outside the medium and  is on
the mass shell.  The mass shell condition can be expressed as
 \begin{eqnarray}
(p-q_1-q_2)^2-m^2~\sim~ -2p\cdot q_1-2p\cdot q_2 ~\sim~ 0.
\label{6}
\end{eqnarray}
The symmetrized sum of the Feynman amplitudes
$M_a$ and $M_b$ in the high-energy limit is
\begin{eqnarray}
\hspace*{-0.3cm}
M
\! \sim \!
\frac{g^4}{2m}  \frac{2 p\cdot {\tilde a}_1}{m_1 q_1^2}
\frac{ 2 p\cdot {\tilde a}_2}{m_2 q_2^2 }
\!\left (\!
 \frac{1}{2 p \cdot  q_1 - i \epsilon} 
\!+\!\frac{1}{2 p \cdot  q_2 - i \epsilon}
\!\right )\!.
\label{15a}
\end{eqnarray} 
Note that the amplitudes $M_a$ and $M_b$ correlate with each other
because of the mass-shell condition (\ref{6}).  The real parts of the
amplitudes destructively cancel, and the imaginary parts interfere and
add constructively, to result in sharp distributions at $p \cdot q_1$
$\sim$ 0 and $p \cdot q_2$ $\sim$ 0,
\begin{eqnarray}
\hspace*{-0.5cm}
M
\!\sim\!
\frac{g^4}{2m}  \frac{2 p\cdot {\tilde a}_1}{m_1q_1^2}
\frac{ 2 p\cdot {\tilde a}_2}{m_2q_2^2 }
\biggl  \{\!
 i \pi \Delta(2 p \cdot  q_1)\!+\!i\pi \Delta(2 p \cdot  q_2)
\!\biggr \}\!,
\label{16a}
\end{eqnarray} 
where
\begin{eqnarray}
\Delta(2p\cdot q_1)=\frac{1}{\pi}\frac{\epsilon}
{(2 p \cdot  q_1)^2+\epsilon^2},\nonumber
\end{eqnarray} 
which approaches the Dirac delta function
$\delta(2p\cdot q_1)$ in the limit $\epsilon \to 0$.

We consider next the case of the (coherent) multiple collisions of a
fast fermion with three fermion scatterers in the reaction
$p+a_a+a_2+a_3 \to p'+a_1'+a_2'+a_3'$.  The Feynman diagrams for the
multiple collisions process are shown in Fig. 2.

\begin{figure} [h]
\includegraphics[scale=0.65]{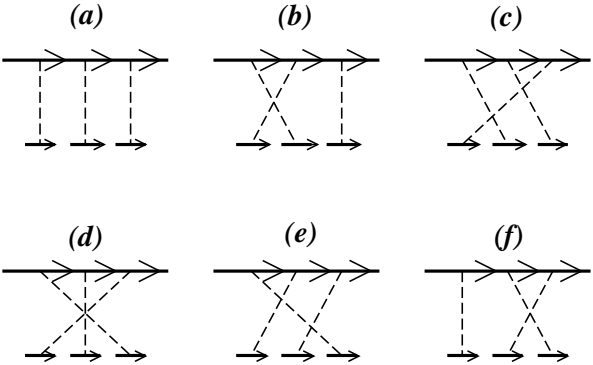}
\caption{ Feynman diagrams for the multiple collisions of a fast
  fermion with three medium fermions in different permutations of the
  exchanged bosons.  }
\end{figure}
 In the high-energy limit, the Feynman amplitude for the collision is 
\begin{eqnarray}
{M}&=&\frac{g^6}{2m} \frac{2 p\cdot {\tilde a}_1}{m_1}\frac{ 2 p\cdot {\tilde a}_2}{m_2}\frac{  2p\cdot {\tilde a}_3 }{m_3} \frac{1}{ q_1^2 q_2^2q_3^2  }
\nonumber\\
 &\times &\biggl \{ \biggl [ \frac{1}{(2p\cdot q_1- i \epsilon)(2p\cdot q_1+2p\cdot q_2- i \epsilon)}
\nonumber\\
& &+ \frac{1}{(2p\cdot q_2- i \epsilon)(2p\cdot q_1+2p\cdot q_2- i \epsilon)}
\biggr ]
\nonumber\\
&+&
\biggl [
\frac{1}{(2p\cdot q_2- i \epsilon)(2p\cdot q_2+2p\cdot q_3- i \epsilon)}
\nonumber\\
& &+ \frac{1}{(2p\cdot q_3- i \epsilon)(2p\cdot q_2+2p\cdot q_3- i \epsilon)}
\biggr ]
\nonumber\\
&+&
\biggl [
\frac{1}{(2p\cdot q_3- i \epsilon)(2p\cdot q_3+2p\cdot q_1- i \epsilon)}
\nonumber\\
& &+ \frac{1}{(2p\cdot q_1- i \epsilon)(2p\cdot q_3+2p\cdot q_1- i \epsilon)}
\biggr ] \biggr \}.
\label{15}
\end{eqnarray}
Each amplitude term in the curly bracket contains a distribution that
is non-zero in various regions of $p$$\cdot$$q_i$.  These Feynman
amplitudes interfere among themselves.  We can write Eq.\ (\ref{15}) in the form
\begin{eqnarray}
\hspace*{-1.0cm}M=\frac{ g^{6}}{2m} 
\left \{ \prod_{i=1}^3 \frac{2 p\cdot {\tilde a}_i }{m_iq_i^2 } \right\} 
\biggl \{ \prod_{j=1}^{2}\frac{1}
{\sum_{i=1}^j 2p\cdot q_i - i \epsilon}
\nonumber\\
~~~~~~+ {\rm symmetric~permutations}
 \biggr \}.
\end{eqnarray}

Generalizing to the case of the coherent collision of a fast fermion
with $n$ fermion scatterers in the process
\begin{eqnarray}
p+a_1+...+a_n \to p' + a_1'+ ...+a_n',
\label{20}
\end{eqnarray}
the total Feynman amplitude is
\begin{eqnarray}
M=\frac{ g^{2n}}{2m} 
\left \{  \prod_{i=1}^n \frac{2 p\cdot {\tilde a}_i }{m_i q_i^2 } \right \}
{\cal M}(q_1,q_2,...,q_n),
\label{10}
\end{eqnarray}
where  ${\cal M}(q_1,q_2,...,q_n)$ is the
sum of  $n!$ amplitudes involving symmetric
permutations of the exchanged bosons given by
\begin{eqnarray}
{\cal M}(q_1,q_2,...,q_n)&=&\prod_{j=1}^{n-1}\frac{1}
{\sum_{i=1}^j 2p\cdot q_i - i \epsilon}
\nonumber\\
&+& {\rm symmetric~ permutations}.
\label{18a}
\end{eqnarray}
The above sum 
involves extensive cancellations.
Remarkably, it can be shown that this sum 
of $n!$ symmetric permutations
turns out to be a product of sharp distributions centered at $2p \cdot q_i \sim 0$
\cite{Che87},
\begin{eqnarray}
\hspace*{-0.6cm}
{\cal M}(q_1,q_2,..,q_n)\Delta (\sum_{i=1}^n \!2p\cdot q_i)
\!=\! 
(2\pi i)^{n-1}
\!\!\prod_{i=1}^n  
\!\! \Delta (2p\cdot q_i),
\label{13}
\end{eqnarray}
which gives
\begin{eqnarray}
\hspace*{-0.6cm}
{\cal M}(q_1,q_2,...,q_n)\!=\! \frac{(2\pi i)^{n-1}}{n}
\! \sum_{i=1}^n \biggl ( 
\prod_{j=1,j\ne i }^n \!\!\!\! \Delta (2p\cdot q_j) \biggr )  .
\label{13a}
\end{eqnarray}
Eqs.\ (\ref{18a})-(\ref{13a})  indicate that for coherent collisions, there is a destructive
interference of the Feynman amplitudes, leading to a cancellation in most regions of the phase
space.  However there is a constructive interference in some other regions of the the phase space, leading to sharp distributions at $2p \cdot q_i \sim 0$.  
The constraint $\Delta (2p \cdot q_i)$ is mathematically the same as  the constrain  $\Delta ( (p-q_i)-m^2)$ in the high-energy limit, which can be interpreted as the intermediate state $p_{i+1}$=$(p-q_i)$ to be on the mass shell.
It should however be kept in mind that the  intermediate state $p_{i+1}=(p-q_i)$ in the constraint  of Eqs.\ (\ref{10}), (\ref{13}), and  (\ref{13a})  is not a single particle but a  quasi-particles that is  
the result of the interference of many amplitudes.

Equations (\ref{10})-(\ref{13a}) for the total Feynman amplitude for a coherent collision as a
product of delta functions of $2p\cdot q_i$ 
are similar to previous results obtained for the emission of many real
photons or gluons in bremsstrahlung, and for the sum of ladder and
cross-ladder amplitudes in the collision of two fermions
\cite{Che87}-\cite{Lam97b}.

\section{ Cross Section for (1+$n$)-body Collision}

We wish to obtain the cross section for the coherent collision of a fast fermion $p$  on
$n$ fermions $\{a_1,a_2,a_3,...,a_n\}$ in reaction (\ref{20}).   While the Feynman rules to construct the Feynman amplitude are well known,  the differential cross section 
for the reaction (\ref{20}) as  a single 
(1+$n$)-body collision process, written in terms of the Feynman amplitude $M$, does not seem to be  available in the literature.  We would like to examine
the cross section rules in some details in this section.

It should be mentioned that the cross section  rules for the (1+$n$)-body collision depend on the Feynman rules for the Feynman amplitude $M$.  We shall use the Feynman rules  of Refs.\ \cite{Che87,Itz80} for $M$,
which adopt the normalization $\bar u u =1$ for the fermion wave function.  A different normalization of the fermion wave function according to $\bar u u=2m$, as given in Ref.\ \cite{Ber82}, will lead to slightly different cross section rules.
  
To be concrete, we use the auxiliary variable $p_i$ with
 $i=1,...,n$ to label  the intermediate state (or quasi-particle) of the incident fermion prior to its exchange of a virtual boson with scatterer $a_i$  (or after its exchange of a virtual gluon with scatterer $a_{i-1}$),  with
the special case  $p_1=p$ and $p_{n+1}=p'$. 
The differential  cross section for the  collision process (\ref{20}) involving  (1+$n$) fermions is given by 
\begin{eqnarray}
d^{n}\sigma &=& \frac{|M|^2(2\pi)^4 \delta^4 
\{p+\sum_{i=1}^n a_i - p'-\sum_{i=1}^n a_i')}{\{\prod_{i=1}^nf_{pi}\} 
\{\prod_{i=2}^n(m/p_{i0}) \}
 \{\prod_{i=2}^n T_i\}}   
\nonumber\\
&\times&
\frac{d^4{ p}'2m}{(2\pi)^3}
\frac{D_{p'}(p')}{ 2 p_{0}' }
\biggl \{ {\prod_{i=1}^n   \frac{d^4 a_i'2m_i}{(2\pi)^3}
\frac{D_i( a_{i}')}{ 2 a_{i0}}} \biggr \},
\label{21}
\end{eqnarray}
where  $M$ is the total Feynman amplitude
as follows from the Feynman rules of Refs.\ \cite{Che87,Note,Itz80}.  
The quantity $f_{pi}$ is the flux factor for  a two-body collision 
 between  $p_i$ and the scatterer $a_i$ of rest mass $m_i$,
\begin{eqnarray}
f_{pi}&=&\frac{4\sqrt{ (p_i\cdot a_{i})^2 - (m ~m_{iT})^2}}{(2m)(2m_i )},
\label{161}
\end{eqnarray}
and $m_{iT}=\sqrt{m_i^2+{\bb
    a}_{iT}^2}$.  

In Eq.\ (\ref{21}) we construct the flux factor  
for the (1+$n$)-body collision  by requiring that it should be the same for all scatterers.
Hence, it is the product  of flux factors from all scatterers,   $\prod_{i=1}^nf_{pi}$.  We also need to require that in this flux factor, the momentum of the 
incident fermion and the scatterer fermions appear in the same power.   Thus, the flux factor need to be linear in the energy of the colliding particles.   But the product of  $\prod_{i=1}^nf_{pi}$ gives $\prod_{i=1}^n p_{i0}$.  It is therefore necessary to multiply $\prod_{i=1}^nf_{pi}$ by the factor $\prod_{i=2}^{n} (m/p_{i0}) $ to get the flux factor for  the (1+$n$)-body collision that is linear in the energy of all colliding particles, resulting in the factors as given in the cross section formula in Eq.\ (\ref{21}). 
  
The quantity $T_i$, with $\{i=2,..n\}$ is the collision time or the mean lifetime of 
the intermediate state (or quasi-particle) $p_i$ of the incident fast particle prior to the incident particle exchanging a virtual boson with $a_i$ (or after the incident particle has already exchanged of a virtual gluon with scatterer $a_{i-1}$).
It arises in the cross section formula because the cross section is proportional to the transition rate per unit time of collision and thus  depends on the lifetime of the intermediate   state  $p_i$ of the  projectile before it exchanges a virtual boson with the scatterer $a_i$.  The precise value of $T_i$ needs not concern us  at this point as it will be canceled later by the delta function that specifies the energy-momentum  constraint of the intermediate state $p_i$, evaluated at the condition of the  constraint of the delta function.

The states of the incident particle $p'$ and the medium scatterer $i'$ after the collision can come in different forms. They can be described by the functions $D_{p'}/2p_{0}'$ and $D_i/2a_{i0}'$.  For example, the final fast particle $p'$ resides outside the medium after the collisions without  interactions.   Its state can be adequately described as being  nearly on the mass shell,  
\begin{eqnarray}
\frac{D_{p'}(p')}{2p_0'} \sim \frac{\delta (p_0' -\sqrt{ {\bb p' }^2 + m^2})}{2p_0'} =\delta (p'^2-m^2).
\end{eqnarray}
It may subsequently break up as jet fragments  in a collimated narrow cone, in accordance with a jet fragmentation function.   We shall examine the states
of the scatterers in the next section.

To show the validity of  Eq.\ (\ref{21}), we note first that it is clearly correct for the case of a two-body collision with $n=1$, as given in Appendix A-3 of Ref.\ \cite{Itz80}.  We would like to test whether it gives the correct result for the case of quasi-elastic collisions where the answer is known.    
Such a case of quasi-elastic collisions is described by the
generalization of ladder diagrams of Fig.\ 1(a), and 2(a), with the incident particle nearly on the mass shell and $q_{i0}\sim 0$.  
The diagram can be effectively cut at the  intermediate fermion lines, resulting in a sequence of $n$ disjoint two-body collision diagrams, and the many-body cross section is just the product of $n$ two-body collisions.

We would like to show that for the above quasi-elastic collision case 
the Feynman amplitude approach with the cross section rules of Eq.\ (\ref{21})
also leads to the known result.

In the high energy limit, the Feynman amplitude for the  
ladder diagrams  of Fig.\ 1(a), and 2(a) can be generalized
for $n$ scatterers as given by  
\begin{eqnarray}
\hspace*{-0.5cm}M
\!\!\sim\!\!
-\frac{g^{2n}}{2m}\left \{   \prod_{i=1}^{n-1} \frac{2 p_{i}\cdot {\tilde a}_{i}}{m_iq_{i}^2}
 \frac{1}{(p_{i+1}^2-m^2)+ i \epsilon} \right \} \frac{2 p_n\cdot {\tilde a}_n}{m_n q_n^2} .
\end{eqnarray} 
For quasi-elastic collisions with $q_{i0}\sim 0$, the incident particle is nearly on the mass shell with $p_{i+1}^2 \sim m^2$.
As a consequence, the propagator represented by $1/(p_{i+1}^2-m^2 + i\epsilon)$ in the Feynman amplitude can be approximated by the pole term
\begin{eqnarray}
\frac{1}{p_{i+1}^2-m^2 + i\epsilon} \to -i\pi \Delta(p_{i+1}^2-m^2).
\end{eqnarray}
The Feynman amplitude becomes
\begin{eqnarray}
M
\sim
\frac{g^{2n}}{2m} \biggl \{  \prod_{i=1}^{n-1} \frac{2 p_i\cdot {\tilde a}_i}{m_i q_i^2}
 i\pi \Delta(p_i^2-m^2)\biggr \} 
 \frac{2 p'\cdot {\tilde a}_n}{m_n q_n^2}.
\end{eqnarray}
We substitute this Feynman amplitude in Eq.\ (\ref{21}), and we obtain
 \begin{eqnarray}
d^{n}\sigma &=& 
\frac{1}
{\{\prod_{i=1}^nf_{pi}\} \{\prod_{i=2}^n(m/p_{i0}) T_i\} } 
\frac{g^{4n}}{(2m)^2}
\nonumber\\
&\times&
 \biggl \{  {\prod_{i=1}^{n-1}  \frac{(2 p_i\cdot {\tilde a}_i)^2}{m_i^2 q_i^4}
 \pi^2 [\Delta(p_i^2-m^2)]^2 \frac{d^4 a_i'2m_i}{(2\pi)^3}
\frac{D_i( a_{i}')}{ 2 a_{i0}}}\biggr \} 
\nonumber\\
&\times&
  \frac{(2 p'\cdot {\tilde a}_n)^2}{m^2q_n^4}
(2\pi)^4 \delta^4 
\{p+\sum_{i=1}^n a_i - p'-\sum_{i=1}^n a_i')
\nonumber\\
&\times&
\frac{d^4{ p}'2m}{(2\pi)^3}
\frac{D_{p'}(p')}{ 2 p_{0}' }
 \frac{d^4 a_n'2m_n}{(2\pi)^3}
\frac{D_n( a_{n}')}{ 2 a_{n0}}.
\label{27}
\end{eqnarray}
On the other hand, for the two-body collision   $p_i+a_i \to p_{i+1} + a_i'$, the Feynman amplitude is
\begin{eqnarray}
M_{pi}^{2B}=\frac{g^2}{2m}\frac{2 p_i\cdot \tilde a_i}{m_i q_i^2}.
\label{28}
\end{eqnarray}
Upon writing $\Delta(p_{i+1}^2-m^2)$ in terms of the mean lifetime $T_{i+1}=1/\Gamma_{i+1}$ of the intermediate state of $p_{i+1}$, we have
\begin{eqnarray}
\Delta(p_{i+1}^2-m^2)
= 
\frac{\Gamma_i/(4\pi p_{(i+1)0})}{ \left [  p_{(i+1)0}-\sqrt{\bb p_{i+1}^2+m^2}\right ]^2 + \Gamma_{i+1}^2/4}\nonumber,
\end{eqnarray}
we get
\begin{eqnarray}
|\Delta(p_{i+1}^2-m^2)|^2
=\Delta(p_{i+1}^2-m^2)\frac{1}{2p_{(i+1)0}}  \frac{2T_{i+1}}{\pi},
\label{30}
\end{eqnarray}
where the mean lifetime  $T_{i+1}$ cancels the corresponding factor in the denominator of Eq.\ (\ref{27}). 
Using Eqs.\ (\ref{28}) and (\ref{30}) in Eq.\ (\ref{27}), we obtain
\begin{eqnarray}
d^{n}\sigma &=&
{\prod_{i=1}^{n-1}\frac{  |M_{pi}^{2B}|^2}{f_{pi}} (2m) 2\pi 
\Delta(p_{i+1}^2-m^2)
\frac{d^4 a_i'2m_i}{(2\pi)^3}
\frac{D_i( a_{i}')}{ 2 a_{i0}}}
\nonumber\\
&\times&
 \frac{ |M_{pn}^{2B}|^2}{f_{pn}} 
(2\pi)^4 \delta^4 
(p'+\sum_{i=1}^n q_i -p)
\nonumber\\
&\times&
\frac{d^4{ p}'2m}{(2\pi)^3}
\frac{D_{p'}(p')}{ 2 p_{0}' }
 \frac{d^4 a_n'2m_n}{(2\pi)^3}
\frac{D_n( a_{n}')}{ 2 a_{n0}}.
\label{33}
 \end{eqnarray}
Inserting the unity factor
$\prod_{i=1}^{n-1} \delta(p_i+a_i-p_{i+1}-a_i') d^4p_{i+1}=1$ into the above equation and noting that
the differential cross section for $p_i+a_i \to p_{i+1} + a_i'$ is
\begin{eqnarray}
d\sigma_{pi}^{2B} &=& \frac{|M_{pi}^{2B}|^2}{f_{pi}}   (2\pi)^4 \delta^4 
(p_{i}+a_i -p_{i+1} - a_i')
\nonumber\\
&\times&
\frac{d^4{ p_{i+1}}2m}{(2\pi)^3}
\frac{D_{i+1}(p_{i+1})}{ 2 p_{(i+1)0} }
\frac{d^4 a_i'2m_i}{(2\pi)^3}
\frac{D_i( a_{i}')}{ 2 a_{i0}},
\label{29}
\end{eqnarray}
we can rewrite Eq.\ (\ref{33}) as 
\begin{eqnarray}
d^{n}\sigma &=& \prod_{i=1}^n d\sigma_{pi}^{2B},
\end{eqnarray}
which is the cross section for quasi-elastic collisions. 
This show that Eq.\ (\ref{21}) gives the correct cross section for the case
for which the answer is known.  It
can therefore be used for  a general treatment for the cross section in terms of many-body Feynman amplitudes
in the general case of coherent collisions.

Equation (\ref{21}) exhibits explicitly the rules for the cross section for 
 coherent collision of an incident fast fermion with $n$  fermion scatterers.  The other cases for incident gluons and gluon scatterers will be taken up in Section IX. 

\section{Consequences of the BE Interference on the Recoil of Fermion Scatterers}

The states of a medium scatterer after the collision can come in different forms.   As it acquires a large amount of energy and momentum from the incident jet, the medium scatterer  
can be far off the mass shell and it may subsequently fragment into a cluster of on-mass-shell particles.  For an incident jet of energy 10 GeV with scatterers in a medium, it is however unlikely that the final scatterers themselves be energetic enough or far off the mass shell  to fragment
as clusters.  Far more likely is the case of the scatterer $a_i'$ to acquire a fraction of the incident jet energy and momentum and become only slightly off the mass shell, with the degree of its being off-mass shell described by a width $\Gamma_i$.  As the medium particles reside in an interacting medium,  it will be subject to the interactions of the medium before and after the collision.    We can describe 
the mass-shell constraint for the 
of medium scatterer $a_i'$  after the collision as
\begin{eqnarray}
D_i( a_{i}') =\frac{\Gamma_i/2\pi}{  [a_{i0}'-
\sqrt{(\bb a_i' - g_i\bb A)^2 + (m_i+S)^2 } +g_i A_0]^2 + \frac{\Gamma_i^2}{4} },\nonumber
\end{eqnarray}
where $A=\{\bb A, A_0\}$ and $S$ are the vector and scalar mean fields
experienced by the medium scatterer $a_i'$ after the collision,
respectively.     As the mean
fields and scatterer widths increase with medium density and are presumably
quite large and dominant for a dense medium, we shall approximately
represent $D_i$ as an average constant that is only a weak function of $a_i'$.  Other descriptions of $D_i/2a_{i0}'$ for the states of the scatterers are also possible but may not be as general; they can be the subjects for future investigations.

We  integrate over 
$\delta(p'+\sum q_i -p) d^4 p'$ and
change variables from $
a_i'$ to $ q_i=a_i'-a_i$.  Writing out the matrix $M$ explicitly as in Eq.\ ({\ref{10}) and using
the relations in Eq.\ (\ref{13}) and (\ref{13a}), we obtain from Eqs.\ (\ref{21})
\begin{eqnarray}
d^{n} \sigma
&&=
  \frac{g^{4n}(2m)^{n-1}}{(2\pi)^{n+1} }
\left \{ \prod_{i=1}^{n}\frac{D_i}{f_{ij}} \right \}
\left \{\prod_{i=1}^{n-1} 
\frac{|\Delta(2 p\cdot q_i)|^2 dq_{i0}}{(m/p_{(i+1)0}) T_{i+1}}
\right \}
\nonumber\\
&&\times
\Delta(2 p\cdot q_n)dq_{n0}
\left \{ \prod_{i=1}^n \!\!
 \frac{(2 p \cdot {\tilde a}_i)^2 }{m^2 2 a_{i0}'}
\frac{d \bb q_{iT} dq_{iz} }{ q_i^4 }\!\right \} \!,
\label{18}
\end{eqnarray}
where for simplicity, we have taken $m_i=m$.
The distribution
$\Delta(p \cdot q_i)$ can be written as
\begin{eqnarray}
\Delta (2p\cdot q_i)=\frac{1}{p_0+p_z}
\Delta  (q_{i0}-q_{iz} - \zeta),
\end{eqnarray}
where $\zeta$ is 
\begin{eqnarray}
\zeta=\frac{ -(p_{0}-p_z)( q_{i0}+q_{iz}) + 2{\bb p}_T\cdot\bb q_T}{p_0+p_z} ,
\end{eqnarray}
and  the quantity $\zeta$ approaches zero in the high-energy limit of large $p_0$.
The function $\Delta  (q_{i0}-q_{iz} - \zeta)$ provides the constraint at
$ q_{i0}-q_{iz} \sim 0$.
As a consequence
of the constraint, the integration
$\Delta  (q_{i0}-q_{iz} - \zeta) dq_{i0}$ can be carried out, yielding
\begin{eqnarray}
\hspace*{-0.6cm}
\int  \!\!\!\Delta (2p\cdot q_i) dq_{i0}
\!=\!\!\int  \!\!\!
\frac{1}{p_z\!+\!p_0}\Delta(q_{i0}\!-\!q_{iz} \!+\! \zeta) dq_{i0}
\!=\!\frac{1}{2p_z}.
\end{eqnarray}
Noting that $\Delta(2p\cdot q_i)$$\sim$$(p$-$q_i)^2$-$m^2$,  we 
can express $\Delta(2p\cdot~q_i)$  in terms of the mean lifetime $T_{i+1}=1/\Gamma_{i+1}$ of the intermediate state 
$p-q_i = p_{i+1}$
of the fast projectile, after it has exchanged of a virtual boson with the scatter $i$, as  
\begin{eqnarray}
\hspace*{-0.5cm}
\Delta (2p\cdot q_i)=
\frac{\Gamma_{i+1}/4\pi p_{(i+1)0}}
{[p_{(i+1)0}-\sqrt{\bb p_{(i+1)}^2 +m^2}]^2+\Gamma_{i+1}^2/4}.
\end{eqnarray}
then we have 
\begin{eqnarray}
| \Delta (2p\cdot q_i)|^2 
=\Delta (2p\cdot q_i)\frac{T_{i+1}}{\pi p_{(i+1)0}}.
\end{eqnarray}
The above mean lifetime $T_{i+1}$ cancels 
the mean lifetime $T_i$ in the denominator of Eq.\ (\ref{18}).
The  boson propagator 
$q_i^2$ in the 
denominator 
 becomes
\begin{eqnarray}
q_i^2=(q_{i0}+q_{iz}) ( q_{i0}-q_{iz})   - |{\bb q}_{iT}|^2\approx - |{\bb q}_{iT}|^2 .
\end{eqnarray}
We obtain 
\begin{eqnarray}
d^n \sigma  &=&\frac{1}{4}
\left (  \frac{\alpha^2}{mp_z} \right )^{ n}
\left \{ \prod_{i=1}^{n}\frac{8D_{i}}{f_{pi}} \frac{ (2p\cdot {\tilde a}_i)^2 }{m  2a_{i0}'} \frac{dq_{iz} d{\bb q}_{iT}}{   |\bb q_T|^4} \!\!\right \}\! ,
\label{22}
\end{eqnarray}
where $\alpha$=$g^2/4\pi$. The last  factor in the curly bracket is dimensionless, and 
$d^n\sigma$ has the dimension of $(\alpha^2/mp_z)^n$, as it should be.

To find the probability distribution for the longitudinal momentum
transfer $q_{iz}$, we introduce the fractional longitudinal momentum
kick
\begin{eqnarray}
x_i=\frac{q_{iz}}{p_z},~~~~~
dq_{iz}=p_z dx_i.
\end{eqnarray}
To investigate the $x_i$ dependence of the factor  $(2p$$\cdot$${\tilde a_i})^2/{2a_{i0}'}$ in Eq.\ (\ref{22}), we note from
Eq.\ (\ref{5}) that ${\tilde a_i}$ can be written as a function of
$q_i$ and $a_i$,
\begin{eqnarray}
\tilde a_i \sim \sqrt{\frac{a_{i0}+m}{{a_{i0}'+m} }}\frac{q_i}{2}
+\frac{a_{i0}'+m + a_{i0}+m}{\sqrt{(a_{i0}'+m)(a_{i0}+m) }}
\frac{a_i}{2}.
\end{eqnarray}
Because of the  $\Delta(2p\cdot q_i)$ constraint,
the factor $(2p$$\cdot$${\tilde a_i})^2/{2a_{i0}'}$  in Eq.\ (\ref{22}) 
becomes 
\begin{eqnarray}
\frac{(2p\cdot {\tilde a_i})^2}{2a_{i0}'}
 \sim  \frac{(a_{i0}'+ a_{i0})^2}{2(a_{i0}')^2 }
\frac{(p\cdot {a_i})^2}{a_{i0}}
\equiv \kappa_i \frac{(p\cdot {a_i})^2}{a_{i0}}.
\end{eqnarray}
We obtain from Eq.\ (\ref{22})
\begin{eqnarray}
d^n \sigma 
&=&
\left \{\frac{1}{4}
\left ( \frac{\alpha^2}{m}\right )^{n}
\left ( \prod_{i=1}^{n}\frac{8D_{i}}{f_{ij}} 
 \frac{\kappa_i (p\cdot a_i)^2}{ma_{i0}} \right )\right \}
\nonumber\\
&\times&
\frac{dx_1 dx_2 ... dx_n d{\bb q}_{1T}d{\bb q}_{2T}...d{\bb q}_{nT}}
{  |\bb q_{1T}|^4~|\bb q_{2T}|^4~...~|\bb q_{nT}|^4}.
\label{25a}
\end{eqnarray}
The fermion scatterers can possess different initial energies $
a_{i0}$ at the moment of their collisions with the energetic jet.  In
the case when $ a_{i0} \ll q_{i0}$, the factor $\kappa_i$ approaches
$1/2+O(a_{i0}/q_{iz})$ with $(a_{i0}/q_{iz}) \ll 1$.  In the other
extreme when $ a_{i0} \gg q_{i0} $, the factor $\kappa$ approaches $2
+ O(q_{iz}/a_{i0})$ with $(q_{iz}/a_{i0}) \ll 1$.  The dependence of
$\kappa_i$ on $x_i$ and $\bb q_{iT}$ is weak in either limits, and such dependencies can be neglected in
our approximate estimate.  We obtain then 
\begin{eqnarray}
\hspace*{-0.5cm}d^n \sigma 
\sim 
C_q
 \frac{dx_1 dx_2 ... dx_n 
d{\bb q}_{1T}  d{\bb q}_{2T} ...d{\bb q}_{nT} }
{  |\bb q_{1T}|^4~|\bb q_{2T}|^4~...~|\bb q_{nT}|^4},
\label{48}
\end{eqnarray}
where $C_q$ is
\begin{eqnarray}
C_q=
\left \{\frac{1}{4}
\left ( \frac{\alpha^2}{m}\right )^{n}
\left ( \prod_{i=1}^{n}\frac{8D_{i}}{f_{ij}} 
 \frac{\kappa_i (p\cdot a_i)^2}{ma_{i0}} \right )\right \},
\end{eqnarray}
the quantity $C_q$ is a weak function of  $x_i$ and
${\bb q}_{iT}$ and can be approximated as a constant.

By symmetry, the fraction of momentum transfers $x_i$
for different scatterers should be approximately the same on the
average, and $x_i^{\rm max} $$\sim$ $1/n$.  Then as far as $x_i$ is
concerned, the average distribution is
\begin{eqnarray}
\frac{dP}{dx_i}  \sim  n~ \Theta(\frac{1}{n}-x_i),
\label{35}
\end{eqnarray}
and the average longitudinal momentum fraction 
is
\begin{eqnarray}
\langle x_i\rangle \sim \frac{1}{2n}~~~{\rm or}~~\langle q_{iz}\rangle \sim \frac{p_z}{2n}.
\label{38}
\end{eqnarray}
The above result in Eq. (\ref{35}) indicate that the probability distribution of the longitudinal momentum transfer of a scatterer is approximately flat in $x_i$.  The quasi-elastic scattering with $q_{iz}\sim0$ occurs, but with about the same probability as other longitudinal momentum transfer up to $q_{iz}=p_z/n$.  On the average, a scatterer acquires approximately $1/2n$ fraction of the incident jet longitudinal momentum. 

Because of the recoils of the scatterers, the incident jet loses its longitudinal momentum.
On the average, the jet loses a longitudinal momentum fraction of $1/2n$ after each collision with a scatterer.
When the jet emerges out of the medium after colliding with $n$ scatterers, the (average) fractional jet longitudinal momentum loss due to scatterers recoils is approximately
\begin{eqnarray}
\hspace*{-0.4cm}\langle {\rm fractional~jet~longitudinal~momentum~loss } \rangle &\sim& n\times \frac{1}{2n}
\nonumber\\
&=& \frac{1}{2}.
\end{eqnarray}

Having obtained the differential cross section Eq.\ (\ref{48}), we can infer the distribution of the scatterers with respect to the incident particle axis.
Equation  (\ref{48}) indicates that the reaction has a high probability for the occurrence of
small values of $ |{\bb q}_{iT}|$,
 in the passage of an energetic fermion making coherent collisions with medium
partons. The singularities at $|\bb q_{iT}|
\sim 0$ in Eq.\ (\ref{48}) correspond to the case of infrared
instabilities that may be renormalized, and a momentum cut-off
$\Lambda_{\rm cut}$ may be introduced.
As $\bb q_{iT}$ are independent degrees of freedom,
Eq.\ (\ref{22}) or (\ref{48}) shows that the standard deviation of the
transverse momentum distribution of the scattered incident particle is related to the transverse momentum transfers to the scatterers by 
\begin{eqnarray}
\langle (\bb p_{T}')^2 \rangle =
\langle (\sum_{i=1}^n  \bb q_{iT})^2 \rangle 
= \sum_{i=1}^n \langle \bb q_{iT}^2 \rangle = n\langle \bb q_{iT}^2 \rangle,
\end{eqnarray}
as in a random walk in the transverse direction.

Equations  (\ref{48})-(\ref{38}) indicate further
that the scatterers acquire an
average longitudinal momentum $\langle q_{iz}\rangle \sim p_z/2n$ that
is expected to be much greater than $\langle |\bb q_{iT}|\rangle$.
Thus, in a coherent collision  there is a collective quantum many-body effect arising from
Bose-Einstein interference such that the fermion scatterers emerge in
the direction of the incident particle, each carrying a fraction of
the forward longitudinal momentum of the incident particle that is
inversely proportional to twice the number of scatterers, $\langle
q_{iz}\rangle \sim {p_z}/{2n}$.

We have presented an explicit derivation of the differential cross section as a function of the momenta of the recoil scatterers by 
making many simplifying assumptions.     The explicit derivation
has the advantage that it  allows future  refinements on  some parts of the calculation by modifying some of the simplifying  assumptions.

\section{Bose-Einstein Interference for  Collisions  in Non-Abelian Theory}

~~~The above considerations for the Abelian theory can be extended to
the non-Abelian theory.  As an example, we consider a quark jet $p$
making coherent collisions with quarks $a_1$ and $a_2$ in the reaction
$p+a_1+a_2 \to p'+a_1'+a_2'$, in the non-Abelian theory.  We shall
neglect four-particle vertices and loops, which are of higher-orders.
The Feynman diagrams are then the same as those in Fig.\ 1.  One
associates each quark vertex with a color matrix
$T_{\alpha,\beta}^{(p,1,2)}$ where the superscript $p,1,$ or 2
identifies the quark $p$, $a_1$, or $a_2$, and the subscripts $\alpha$
or $\beta$ give the $SU(3)$ color matrix index.  The Feynman amplitude
$M_a$ for diagram 1(a) is
\begin{eqnarray}
M_a &=& -g^4 {\bar u}({\bb p}')
 T_\beta^{(p)} \gamma_\nu \frac{1}{\slashed p - \slashed q_1-m+ i \epsilon'} 
T_\alpha^{(p)}\gamma_\mu u({\bb p})
\nonumber\\
&\times&
\frac{1}{q_2^2}  {\bar u}({\bb a}_2')
T_\beta^{(2)} \gamma_\nu u({\bb a}_2)
\frac{1}{q_1^2}  {\bar u}({\bb a_1}')
T_\alpha^{(1)}\gamma_\mu u({\bb a}_1).
\nonumber
\end{eqnarray}
The Feynman  amplitude $M_b$ for  diagram 1(b) is
\begin{eqnarray}
M_b& =& -g^4 {\bar u}({\bb p}')
T_\alpha^{(p)}  \gamma_\mu \frac{1}{\slashed p - \slashed q_2-m+ i \epsilon'} 
T_\beta^{(p)} \gamma_\nu u({\bb p})
\nonumber\\
&\times& 
\frac{1}{q_2^2}  {\bar u}({\bb a}_2')
T_\beta^{(2)} \gamma_\nu u({\bb a}_2)
\frac{1}{q_1^2}  {\bar u}({\bb a_1}')
T_\alpha^{(1)}\gamma_\mu u({\bb a}_1).\nonumber
\end{eqnarray}
In the high-energy limit for coherent collisions, the sum of the Feynman amplitudes is
\begin{eqnarray}
M
&\sim&
\frac{g^4}{2m}  \frac{2 p\cdot {\tilde a}_1}{m}
\frac{ 2 p\cdot {\tilde a}_2}{m}
\frac{1}{q_2^2 q_1^2}
\nonumber\\
&\times&
\left (\frac{T_\beta^{(p)} T_\alpha^{(p)}T_\beta^{(2)} T_\alpha^{(1)}}{2p\cdot q_1 -i\epsilon}
+\frac{ T_\alpha^{(p)} T_\beta^{(p)}T_\beta^{(2)} T_\alpha^{(1)}}{2p\cdot q_2 -i\epsilon}
\right )\!\!.
\label{44}
\end{eqnarray}
We can rewrite the product of the color matrices for the  quark jet $p$ as 
\begin{eqnarray}
T_\beta^{(p)} T_\alpha^{(p)}=\frac{1}{2}\left ( [ T_\beta^{(p)}, T_\alpha^{(p)}]_+ + [ T_\beta^{(p)}, T_\alpha^{(p)}]_- \right ), \\
T_\alpha^{(p)} T_\beta^{(p)}=\frac{1}{2}\left ( [ T_\beta^{(p)}, T_\alpha^{(p)}]_+ - [ T_\beta^{(p)}, T_\alpha^{(p)}]_- \right ) .
\end{eqnarray}
We label the propagators in Eq.\ (\ref{44})
as   ${\cal M}_{a}$ and ${\cal M}_{b}$,
\begin{eqnarray}
{\cal M}_a = \frac{1}{2p\cdot q_1 -i\epsilon},
~~{\rm and~~}
{\cal M}_b = \frac{1}{2p\cdot q_2 -i\epsilon}.
\end{eqnarray}
The Feynman amplitude for coherent collision  is then
\begin{eqnarray}
M&&
\sim
\frac{g^4}{2m}  \frac{2 p\cdot {\tilde a}_1}{m}
\frac{ 2 p\cdot {\tilde a}_2}{m}
\frac{1}{q_2^2 q_1^2}
\nonumber\\
&&\times
\biggl \{({\cal M}_a+{\cal M}_b) 
\frac{[T_\beta^{(p)},T_\alpha^{(p)}]_+T_\beta^{(2)} T_\alpha^{(1)}}{2}
\nonumber\\
&&~~+
 ({\cal M}_a-{\cal M}_b) 
\frac{[T_\beta^{(p)},T_\alpha^{(p)}]_- T_\beta^{(2)} T_\alpha^{(1)}}{2}
 \biggr \}. \label{49}
 \end{eqnarray}
The first term inside the curly bracket has the same space-time
structure as what one obtains in the Abelian theory.  It is given by
the Abelian Feynman amplitude in Section II, multiplied by the color
factor 
\begin{eqnarray}
C_{{}_{CF}}=\frac{T_\beta^{(2)} T_\alpha^{(1)}}{2}
[T_\beta^{(p)},T_\alpha^{(p)}]_+ .
\end{eqnarray}
The sum of
${\cal M}_a$ and ${\cal M}_b$ leads to sharp distributions at $p\cdot
q_1\sim 0$ and $p\cdot q_2\sim 0$,
\begin{eqnarray}
{\cal M}_a  + {\cal M}_b
=i \Delta(2 p \cdot  q_1)+i\Delta(2 p \cdot  q_2).
\end{eqnarray} 
The second term is new and occurs only in the non-Abelian theory, as
it involves the commutator of $T_b^{(p)}$ and $T_a^{(p)}$.  It
involves the difference of ${\cal M}_a$ and ${\cal M}_b$ in which the
sharp distributions cancel each other, leaving a broad distribution,
\begin{eqnarray}
{\cal M}_a - {\cal M}_b  &=&
\frac{4 p \cdot  q_1}
{(2 p \cdot  q_1)^2+\epsilon^2}.
\end{eqnarray}

From the above analysis, we find that the color degrees of freedom in
QCD bring in additional properties to the Feynman amplitude.
Bose-Einstein symmetry with respect to the interchange of gluons in
QCD involves not only the space-time exchange symmetry but also color
index exchange symmetry.  The total exchange symmetry can be attained
with symmetric space-time amplitudes and symmetric color index factors
as in the first ${\cal M}_a + {\cal M}_b$ term in Eq.\ (\ref{49}).
The total symmetry can also be attained with space-time antisymmetry
and color index antisymmetry, as in the second ${\cal M}_a-{\cal M}_b$
term in Eq.\ (\ref{49}) .

 We consider next the case for the collision of a quark jet with
 three quark scatterers in the reaction $p+a_a+a_2+a_3 \to
 p'+a_1'+a_2'+a_3'$ in the non-Abelian theory.  The Feynman diagrams
 for the collision process are the same as those in Fig. 2 where we
 associate the color matrices $T_a^{(1)}$, $T_b^{(2)}$, $T_c^{(3)}$ of
 color indices $a, b, c$ with fermion scatterers $a_1$, $a_2$ and
 $a_3$ respectively.  In the high-energy limit, the Feynman amplitude
 for the collision is
\begin{eqnarray}
&M&=\frac{g^6}{2m} \frac{2 p\cdot {\tilde a}_1 2 p\cdot {\tilde a}_2 2 p\cdot {\tilde a}_3 }{m^3 q_1^2 q_2^2q_3^2  }T_a^{(1)} T_b^{(2)} T_c ^{(3)}
\nonumber\\
 &\times &\biggl \{ 
T_a T_b T_c M_{123}+ T_a T_c T_b M_{132}
   +T_b T_c T_a M_{231}
\nonumber\\
& &+ T_b T_a T_c M_{213}
   +T_c T_a T_b M_{312}+ T_c T_b T_a M_{321}
\biggr \},
\label{51}
\end{eqnarray}
where $T_{a,b,c}$ without a superscript are the color matrices for the
incident quark jet $p$, and the amplitudes
$M_{123},M_{132},M_{231},M_{213},M_{312},M_{321}$ are sequentially the
six terms in the curly brackets of the Abelian Feynman amplitudes in
Eq.\ (\ref {15}).  It is easy to show that the quantity in the curly
bracket of Eq.\ (\ref{51}) can be re-written as
\begin{eqnarray}
{\cal M}&=&
 \left \{ T_a [T_b,T_c]_+ + T_b  [T_c, T_a]_++T_c  [T_a, T_b]_+ \right \}
\nonumber\\
& &  
\times[ M_{123}+M_{132}
+ M_{231}+M_{213}
+ M_{312}+M_{321}     ] / 2
\nonumber\\
&-& T_a [T_b, T_c]_+[ M_{231}+M_{213}
+ M_{312}+M_{321} ]/2
\nonumber\\
&-&   T_b  [T_c, T_a]_+[M_{123}+M_{132}
+M_{312}+M_{321}]/2
\nonumber\\
&-& T_c  [T_a, T_b]_+[M_{123}+M_{132}
+ M_{231}+M_{213} ]/2
\nonumber\\
&+&   T_a [T_b, T_c]_- \left (M_{123}-M_{132}\right)/2
\nonumber\\
&+&   T_b  [T_c, T_a]_-\left ( M_{231}-M_{213}\right )/2
\nonumber\\
&+&   T_c  [T_a, T_b]_-\left ( M_{312}-M_{321}\right)/2.
\label{53}
\end{eqnarray}

The first term on the right hand side involves the symmetric sum of
the space-time part of permuted amplitudes as in the Abelian case,
multiplied by the symmetric permutation of the color indices.
It yields a Feynman amplitude that is just the Abelian Feynman
amplitude multiplied by the color factor
\begin{eqnarray}
C_{{}_{CF}}
\!\!=  \! \frac{T_a^{(1)} T_b^{(2)} T_c ^{(3)} }{2}\!\biggl \{\! T_a [T_b,\!T_c]_+ \!+ \!T_b  [T_c,\!T_a]_+\!+\!T_c  [T_a,\!T_b]_+ \!\!\biggr \}\!.
\nonumber
\end{eqnarray}
The other terms involve partial symmetry and antisymmetry with respect
to the exchange of color indices.  Similar studies on the collision of a
jet with $n$ partons can be carried out as in
\cite{Che69,Che87,Fen96,Fen97,Lam96a,Lam97,Lam97a,Lam97b}.  For our
present work, it suffices to note that there will always be a
component of the Feynman amplitude that is symmetric under both
space-time exchange and color index exchange involving the sum of all
space-time amplitude components, similar to the ${\cal M}_a + {\cal
  M}_b$ sum in Eq.\ (\ref{49}) and the first term on the right-hand
side of Eq.\ (\ref{53}).  There will also be other space-time
antisymmetric and color index exchange antisymmetric components.

For the space-time symmetric and color index exchange symmetric
component, the Feynman amplitude is equal to the Abelian Feynman
amplitude multiplied by a color factor. It will exhibit the same
degree of Bose-Einstein interference as in the Abelian theory.
Previous analysis on the longitudinal momentum transfer of recoiling
fermions in the Abelian theory in Section 2 can be applied for the
non-Abelian theory for this space-time symmetric and color index
exchange symmetric component.  There is thus a finite probability for
the presence of delta function constraints to lead to recoiling
quarks receiving significant moment kicks along the direction of the
incident quark jet.

\section{Collision of a Gluon Jet with Quark Scatterers}

It is of interest to generalize the above considerations to the
coherent multiple collisions of a gluon jet.  We shall neglect
four-particle vertices and loops, which are of higher-orders.  The
Feynman diagrams for the collision of a gluon jet with medium quarks
or medium gluons then have structures and momentum flows the same as
those in the collision of a quark jet with quark scatterers.  In the
high-energy limit, the propagators and the three-particle vertices
have approximately the same momentum dependencies, and the
Bose-Einstein symmetry with respect to the interchange of the virtual
bosons is the same.  One expects that aside from the presence of color
factors and color indices, the results for the Bose-Einstein
interference in collisions with a gluon jet or a quark jet should be
similar.  This is so because the high-energy processes are insensitive
to the spins of the colliding particles, as the current carried by a
high energy particle is dominated by its center-of-mass motion, much
more so than its spin current \cite{Fen97,Lo76,Che81}.

It is instructive to study the coherent collision of a fast gluon $p$ with
two quarks in non-Abelian gauge field theories in the reaction
$p+a_1+a_2 \to p' + a_1' +a_2'$ as an examples of the type of BE
interference for the collision of a gluon jet.  The collision process
is represented by the two Feynman diagrams in Fig.\ 3.

\begin{figure} [h]
\includegraphics[scale=0.65]{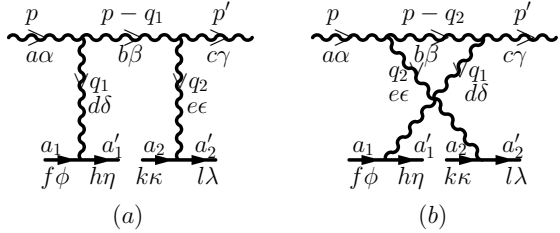}
\vspace*{-0.2cm}
\caption{ Feynman diagrams for the collision of a gluon jet $p$ with
  quark partons $a_1$, $a_2$ with the momentum transfers $q_1$ and
  $q_2$ in the reaction $p+a_1+a_1 \to p'+a_1'+a_2'$.  Here, $\{
  a,b,c,...,k,l\}$ label the color states and $\{\alpha,
  \beta,...,\eta,\lambda\}$ label the momentum components of various
  particles.  }
\end{figure}

We consider the Feynman amplitude matrix element between the
initial three particle state $|p a \alpha, a_1 f \phi, a_2 k \kappa
\rangle$ and the final state $|p' c \gamma, a_1' h \eta, a_2' l
\lambda\rangle$ where $\{ a,f, k,c,h,l \}$ label the color states and
$\{\alpha, \phi,\kappa,\gamma,\eta,\lambda\}$ label components of
various particle currents.  Using the Feynman rules and the overall
phase factors as given in Ref.\ \cite{Che87}, the Feynman amplitude
matrix element of Fig.\ 3(a) is given by
\begin{eqnarray}
&&\langle p' c \gamma, a_1' h \eta, a_2' l \lambda~ | ~M_a~|~p a \alpha,
  a_1 f \phi, a_2 k \kappa \rangle \nonumber\\ &&= g^4\frac{ f_{bec}
    f_{bad}}{q_1^2 q_2^2 (2p\cdot q_1-i\epsilon)} (T^{(2)}_e)_{lk}
  (T^{(1)}_d)_{hf} \epsilon_\gamma (p') \nonumber\\ && \times \{
  g_{\gamma\beta} (p'+p-q_1)_\epsilon +g_{ \beta \epsilon}( -p+q_1-
  q_2)_\gamma +g_{\epsilon \gamma}(q_2-p')_\beta \}
  \nonumber\\ &&\times \{ g_{\beta\alpha} (2p-q_1)_\delta +g_{ \alpha
    \delta}( -p - q_1)_\beta +g_{\delta \beta}(2q_1-p)_\alpha \}
  \epsilon_\alpha^* (p) \nonumber\\ && \times {\bar u}({\bb a}_2')
  \gamma_\epsilon u({\bb a}_2) {\bar u}({\bb a_1}') \gamma_\delta
  u({\bb a}_1),
\label{55}
\end{eqnarray}
where $\epsilon (p)$ and $\epsilon (p')$ are the polarization vectors
for gluons $p$ or $p'$ respectively.  In the high-energy limit in
which $ |{\bb p}| \gg \{ |{\bb q}_i|, |{\bb a}_i|, m \}$, terms of
order $ \{ |{\bb q}_i|/|\bb p|, |{\bb a}_i|/|\bb p|, m/|\bb p| \}$ in
the three-gluon vertices can be neglected, and the helicities can be
assumed to conserve.  We then obtain
\begin{eqnarray}
&&\langle p' c \gamma, a_1' h \eta, a_2' l \lambda ~ |  ~M_a~|~p a \alpha, a_1  f \phi, a_2 k \kappa \rangle
\nonumber\\ 
&&
\!\!\!
= g^4 \frac{(T_e^{(p)})_{cb} (T_d^{(p)})_{ba}
 (T^{(2)}_e)_{lk}
 (T^{(1)}_d)_{hf}}
{q_1^2 q_2^2 (2 p \cdot q_1 - i \epsilon)}
\frac{2p \cdot{\tilde a}_{1}}{m} \frac{2p\cdot {\tilde a}_{2}}{m},
\nonumber \\
\label{56}
\end{eqnarray}
where the coefficients $f_{bec}$ and $f_{bad}$ in Eq.\ (\ref{55}) have
been expressed as matrix elements of matrices $T_d^{(p)}$ and
$T_e^{(p)}$ of the incident gluon $p$ (or $p'$) between gluon color
states \cite{Che87},
\begin{eqnarray}
(T_d^{(p)})_{ba}=if_{dba},\\
(T_e^{(p)})_{cb}=if_{ecb}.
\end{eqnarray}
 Equation (\ref{56})  can be re-written  in a matrix form as
\begin{eqnarray}
&&M_a=
g^4 \frac{2 p\cdot {\tilde a}_1 2 p\cdot {\tilde a}_2}{m^2 q_1^2q_2^2}\frac{T_e^{(p)} T_d^{(p)}T_e^{(2)} T_d^{(1)}}
{2p\cdot q_1-i\epsilon}.
\end{eqnarray}
We can obtain a similar result for the Feynman amplitude for diagram
3(b) by permuting the vertices of the exchange bosons.  As a
consequence, the sum of the two Feynman amplitudes from Figs.\ 3(a)
and 3(b) is given by
\begin{eqnarray}
&&M=
{g^4} \frac{2 p\cdot {\tilde a}_1 2 p\cdot {\tilde a}_2}{m^2 q_1^2q_2^2}\nonumber\\
&&\times  \biggl \{
\frac{ T_e^{(p)} T_d^{(p)}T_e^{(2)} T_d^{(1)}}
{2p\cdot q_1-i\epsilon}
+ \frac{T_d^{(p)} T_e^{(p)}T_e^{(2)} T_d^{(1)}}
{2p\cdot q_2-i\epsilon}\biggr \}.
\label{60}
\end{eqnarray}
We note that the above equation for the collision of a gluon jet is in
the same form as Eq.\ (\ref{44}) for the collision of a quark jet
except with the modification that in the above equation for a gluon
jet, the operator $T^{(p)}$ has matrix elements between gluon color
states whereas the operator $T^{(p)}$ in Eq.\ (\ref{44}) for the quark
jet has matrix elements between quark color states.  As in
Eq.\ (\ref{44}), we can likewise express the product of the color
matrices as
\begin{eqnarray}
T_b^{(p)} T_a^{(p)}=\frac{1}{2}\left ( [ T_b^{(p)}, T_a^{(p)}]_+ + [ T_b^{(p)}, T_a^{(p)}]_- \right ), \\
T_a^{(p)} T_b^{(p)}=\frac{1}{2}\left ( [ T_b^{(p)}, T_a^{(p)}]_+ - [ T_b^{(p)}, T_a^{(p)}]_- \right ) .
\end{eqnarray}
The Feynman amplitude for a gluon jet is then
 \begin{eqnarray}
M
& \sim &
{g^4} \frac{2 p\cdot {\tilde a}_1  2p\cdot {\tilde a}_2}{m^2 q_1^2q_2^2}
\nonumber\\
&\times&
\biggl  \{({\cal M}_a+{\cal M}_b) 
\frac{[T_b^{(p)},T_a^{(p)}]_+T_b^{(2)} T_a^{(1)}}{2}
\nonumber\\
&  &+
({\cal M}_a-{\cal M}_b) 
\frac{[T_b^{(p)},T_a^{(p)}]_- T_b^{(2)} T_a^{(1)}}{2} \biggr \}. 
\end{eqnarray}
Therefore, in the collision of both a gluon or a quark jet with
quarks, there will always be a component of the Feynman amplitude that
is symmetric under both space-time exchange and color index exchange,
involving the sum of all space-time amplitude components.  The
similarity is so close that previous results concerning a quark
jet in collision with quark scatterers apply equally well to a gluon
jet.

\section{Collision of a gluon jet with gluon scatterers}

There is however a small difference in the coherent collisions of a
jet with gluon scatterers.  We can consider the collision of a fast
gluon $p$ with two gluon scatterers $a_1$ and $a_2$ as shown in
Fig.\ 4.

The matrix element of the Feynman amplitude in Fig.\ 4(a) is given by
\begin{eqnarray}
&&\langle p' c \gamma, a_1' h \eta, a_2' l \lambda ~|  ~M_a~|~p a \alpha, a_1  f \phi, a_2 k \kappa \rangle
\nonumber\\ 
&&=
\frac{ g^4 f_{bec}  f_{bad}  f_{kle} f_{fhd}}
{q_1^2 q_2^2 (2 p \cdot q_1 - i \epsilon)}
\epsilon_\gamma (p') \epsilon_\alpha^* (p)\epsilon_\lambda (a_2')\epsilon_\kappa ^*(a_2)\epsilon_\eta (a_1')  \epsilon_\phi^* (a_1)
\nonumber\\
&&
\times \{ g_{\gamma\beta} (p'+p-q_1)_\epsilon
+g_{  \beta \epsilon}( -p+q_1- q_2)_\gamma +g_{\epsilon \gamma}(q_2-p')_\beta   \}
\nonumber\\
&&\times  \{ g_{\beta\alpha} (2p-q_1)_\delta
+g_{  \alpha \delta}( -p - q_1)_\beta +g_{\delta \beta}(2q_1-p)_\alpha  \} 
\nonumber\\
&&\times 
  \{ g_{\kappa\lambda} (-a_2-a_2')_\epsilon
+g_{  \lambda \epsilon}( a_2' + q_2)_\kappa +g_{\epsilon \kappa}(-q_2+a_2)_\lambda  \}
\nonumber\\
&&\times
 \{ g_{\phi\eta} (-a_1 -a_1')_\delta
+g_{  \eta \delta}( a_1' +q_1)_\phi +g_{\delta \phi}(-q_1+a_1)_\eta  \}.
\end{eqnarray}
In the high-energy limit in which  
$|{\bb p}| \gg \{ |{\bb q}_i|, 
|{\bb a}_i|, m\}$  and the helicities can be assumed to conserve, we obtain
\begin{eqnarray}
&&\langle p' c \gamma, a_1' h \eta, a_2' l \lambda ~|  ~M_a~|~p a \alpha, a_1  f \phi, a_2 k \kappa \rangle
\nonumber\\ 
&&
\!\!\!
= {g^4}\frac{(T_e^{(p)})_{cb} (T_d^{(p)})_{ba}
 (T^{(2)}_e)_{lk}
 (T^{(1)}_d)_{hf}}
{q_1^2 q_2^2 (2 p \cdot q_1 - i \epsilon)}
{{4p \cdot{\bar a}_{1}}}
{ {4p\cdot {\bar a}_{2}}}
\nonumber\\
\end{eqnarray}
where 
\begin{eqnarray}
\bar a_i=\frac{a_i+a_i'}{2}.
\label{66}
\end{eqnarray}
We can obtain a similar result for the Feynman amplitude for diagram
4(b) by permuting the vertices of the exchange bosons.  As a
consequence, the sum of the two Feynman amplitudes from Figs.\ 4(a)
and 4(b) is given by
\begin{eqnarray}
&&M=
{g^4} \frac{4 p\cdot {\bar a}_1 4 p\cdot {\bar a}_2}{  q_1^2q_2^2}\nonumber\\
&&\times  \biggl \{
\frac{ T_e^{(p)} T_d^{(p)}T_e^{(2)} T_d^{(1)}}
{2p\cdot q_1-i\epsilon}
+ \frac{T_d^{(p)} T_e^{(p)}T_e^{(2)} T_d^{(1)}}
{2p\cdot q_2-i\epsilon}\biggr \}.
\label{67}
\end{eqnarray}

\begin{figure} [h]
\includegraphics[scale=0.65]{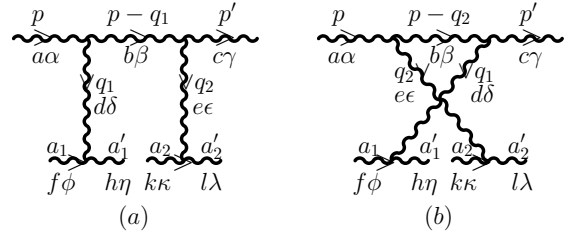}
\vspace*{-0.2cm}
\caption{ Feynman diagrams for the interaction of a gluon jet $p$ with
  gluon scatterers $a_1$, $a_2$ with the momentum transfers $q_1$ and
  $q_2$.  Here, $\{ a,b,c,...,k,l\}$ label the color states and
  $\{\alpha, \beta,...,\eta,\lambda\}$ label the momentum components
  of various particles.  }
\end{figure}

By comparing the results for a quark jet on quark scatterers
[Eq.\ (\ref{44})] with those for a gluon jet on quark scatterers
[Eq.\ {\ref{60})] or on gluon scatterers [Eq.\ (\ref{67})], we obtain
  the following simple rules to obtain from the results of quark jet
  on quark scatterers of Eq.\ (\ref{44}) to those involving gluons:
  (1) provide an overall multiplicative factor of $2m$ when a quark is
  replaced by a gluon, (2) keep the same form of the color operators,
  with the color operators that operate on the quark color states
  switched to operate on the gluon color states, and (3) change the
  momentum variable $\tilde a_i$ of Eq.\ (\ref{5}) for a quark
  scatterer to $\bar a_i$ of Eq.\ (\ref{66}) for a gluon scatterer
  when a quark scatterer is replaced by a gluon scatterer.  The above
  Rule (1) and (2) are well known results that were obtained
  previously on page 198 of Ref.\ \cite{Che87}.  Rule (3) is for our
  case in the collision with quark or gluon scatterers.

For completeness, we can use these rules to obtain the Feynman
amplitude for the collision of a quark jet with two gluon scatterers
as
\begin{eqnarray}
M&=&
\frac{g^4}{2m}  \frac{4 p\cdot {\bar a}_1 4 p\cdot {\bar a}_2}{ q_1^2q_2^2}
\nonumber\\
&\times & 
\biggl \{
\frac{ T_e^{(p)} T_d^{(p)}T_e^{(2)} T_d^{(1)}}
{2p\cdot q_1-i\epsilon}
+ \frac{T_d^{(p)} T_e^{(p)}T_e^{(2)} T_d^{(1)}}
{2p\cdot q_2-i\epsilon}\biggr \}.
\end{eqnarray}

\section{Longitudinal momentum transfer for gluons scatterers}

The cross section rules for Eq.\ (\ref{27}) has been written for the collision of involving fermions (or quarks).
The cross section rule involving  incident gluons and/or gluon scatterers can be generalized by the rule that in replacing a fermion with a gluon, the $2m_{\rm fermion}$ factor for a fermion  is replaced by unity for a gluon.    Thus, we have explicitly
the cross section formula for the coherent collision of an incident gluon with $n$ fermion scatterers  given by
\begin{eqnarray}
d^{n}\sigma_{g+nf} &=& \frac{|M|^2(2\pi)^4 \delta^4 
\{p+\sum_{i=1}^n a_i - p'-\sum_{i=1}^n a_i')}{\{\prod_{i=1}^nf_{pi}\} 
\{\prod_{i=2}^n(1/2p_{i0}) \}
 \{\prod_{i=2}^n T_i\}}   
\nonumber\\
&\times&
\frac{d^4{ p}'}{(2\pi)^3}
\frac{D_{p'}(p')}{ 2 p_{0}' }
\biggl \{ {\prod_{i=1}^n   \frac{d^4 a_i'2m_i}{(2\pi)^3}
\frac{D_i( a_{i}')}{ 2 a_{i0}}} \biggr \},
\end{eqnarray}
where
\begin{eqnarray}
f_{pi}({\rm gluon ~on~fermion})&=&\frac{4 p\cdot a_{i}}{2m_i}.
\end{eqnarray}
Similarly, the cross section formula for the coherent collision of an incident gluon with $n$ gluon scatterers is given by
\begin{eqnarray}
d^{n}\sigma_{g+ng} &=& \frac{|M|^2(2\pi)^4 \delta^4 
\{p+\sum_{i=1}^n a_i - p'-\sum_{i=1}^n a_i')}{\{\prod_{i=1}^nf_{pi}\} 
\{\prod_{i=2}^n(1/2p_{i0}) \}
 \{\prod_{i=2}^n T_i\}}   
\nonumber\\
&\times&
\frac{d^4{ p}'}{(2\pi)^3}
\frac{D_{p'}(p')}{ 2 p_{0}' }
\biggl \{ {\prod_{i=1}^n   \frac{d^4 a_i'}{(2\pi)^3}
\frac{D_i( a_{i}')}{ 2 a_{i0}}} \biggr \},
\end{eqnarray}
where
\begin{eqnarray}
f_{pi}({\rm gluon ~on~gluon})&=&{4 p\cdot a_{i}}.
\end{eqnarray}
The cross section formula for the coherent collision of an incident quark of rest mass $m$ with $n$ gluon scatterers is given by
\begin{eqnarray}
d^{n}\sigma_{g+ng} &=& \frac{|M|^2(2\pi)^4 \delta^4 
\{p+\sum_{i=1}^n a_i - p'-\sum_{i=1}^n a_i')}{\{\prod_{i=1}^nf_{pi}\} 
\{\prod_{i=2}^n(m/p_{i0}) \}
 \{\prod_{i=2}^n T_i\}}   
\nonumber\\
&\times&
\frac{d^4{ p}'}{(2\pi)^3}
\frac{D_{p'}(p')2m}{ 2 p_{0}' }
\biggl \{ {\prod_{i=1}^n   \frac{d^4 a_i'}{(2\pi)^3}
\frac{D_i( a_{i}')}{ 2 a_{i0}}} \biggr \},
\end{eqnarray}
where
\begin{eqnarray}
f_{pi}({\rm fermion ~on~gluon})&=&\frac{4 p\cdot a_{i}}{2m}.
\end{eqnarray}

The results in the last section indicate that for the Feynman amplitude there is a modification
from $\tilde a$ of Eq.\ (\ref{5}) for a
  quark scatterer to $\bar a_i$ of Eq.\ (\ref{66}) for a gluon
  scatterer,  when a quark scatterer is replaced by a
gluon scatterer.  Such a modification brings with it a change in the
longitudinal momentum distribution of the gluon scatterers which we
shall examine in this section.

We consider the coherent multiple collisions of a gluon jet or a gluon
jet on $n$ gluons scatterers.  From the results in Eq.\ (76), the
cross section for the scattering in the space-time symmetric and color
symmetric state is 
\begin{eqnarray}
\hspace*{-0.4cm}
d^n \!\sigma_{g,ng}\!\! &=&\!
\frac{C_{{}_{CF}}}{4}
\biggl ( \!\frac{2\alpha^2}{p_z}
\! \biggr )^{\!n} 
\!\!\left \{ \! \prod_{i=1}^{n} \! \frac{16D_{i}}{f_{pi}} \frac{(2p\cdot \bar a_i)^2} { 2a_{i0}'}
  \frac{dq_{iz} d{\bb q}_{iT}}{   |\bb q_T|^4}
\!\!\right \}\!\! ,
\label{82}
\end{eqnarray}
where $C_{{}_{CF}}$ is the color factor for  the space-time symmetric and color
symmetric component and 
\begin{eqnarray}
f_{pi}=4 p_i\cdot a_{i}.
\end{eqnarray}
 To investigate the longitudinal distribution of
the scatterers, we need to investigate the $q_{iz}$ dependence of the
factor $(2p\cdot {\bar a_i})^2$ in Eq.\ (\ref{82}), which can be
written as
\begin{eqnarray}
{(2p\cdot {\bar a_i})^2}
={[2p\cdot (2a_i+q_i)]^2}.
\end{eqnarray}
Because of the  $\Delta(2p\cdot q_i)$ constraint, the above result gives 
\begin{eqnarray}
{(2p\cdot {\bar a_i})^2} \Delta(2p\cdot q_i)
\sim
{(4p\cdot {a_i})^2}\Delta(2p\cdot q_i),
\end{eqnarray}
where the factor $(4p\cdot {a_i})^2$ is independent of the longitudinal recoil $q_{iz}$ of the scatterer. 
It is convenient to use the transfer rapidity $\xi_i$ to represent
the longitudinal momentum transfer $q_{iz}$,
\begin{eqnarray}
q_{iz}&=&m_{gT} \sinh \xi_i,
\label{66a}\\
 a_{i0}'&=&\sqrt{m_{gT}^2+(a_{iz}+m_{gT}
  \sinh \xi_i)^2},\label{67a}
\end{eqnarray}
where for simplicity the final transverse masses of the scatterers are
taken to be their average value $m_{gT}$.  To make the problem simple,
we can take $a_{iz}$ to have its average value $\langle a_{iz}\rangle$,
which is zero in the medium center-of-momentum frame.
In that approximation, we obtain from Eqs.\ (\ref{66a}) and (\ref{67a})
\begin{eqnarray}
\frac{dq_{iz}}{a_{i0}'} \sim d\xi_i.
\end{eqnarray}
From Eq.\ (\ref{82}),  the cross section becomes   becomes
\begin{eqnarray}
\hspace*{-0.5cm}d^n \sigma 
\sim 
C_g
 \frac{d\xi_1 d\xi_2 ... d\xi_n 
d{\bb q}_{1T}  d{\bb q}_{2T} ...d{\bb q}_{nT} }
{  |\bb q_{1T}|^4~|\bb q_{2T}|^4~...~|\bb q_{nT}|^4},
\end{eqnarray}
where $C_g$ is
\begin{eqnarray}
C_g&=&
\frac{C_{{}_{CF}}}{4}
\biggl ( \!\frac{2\alpha^2}{p_z}
\! \biggr )^{\!n} 
\left \{ \prod_{i=1}^{n}  \frac{16D_{i}}{f_{pi}} \frac{(2p\cdot \bar a_i)^2} { 2}
\right \}. 
\end{eqnarray}
The quantity $C_g$ is a weak function of  $\xi_i$ and
${\bb q}_{iT}$ and can be approximated as a constant. 
The probability distribution is then a flat 
function of $\xi_i$.
  The transfer   
rapidly $\xi_i$ has the upper limit 
\begin{eqnarray}
\xi_{i{\rm max}}=\cosh^{-1}\left \{ \frac{p_0}{nm_{gT}}\right \}.
\end{eqnarray}
 Then as far as $\xi_i$ is
concerned, the average distribution is
\begin{eqnarray}
\frac{dP}{d\xi_i}  \sim  \frac{1}{\xi_{i{\rm max}}}~ \Theta(\xi_{i{\rm max}}-\xi_i).
\end{eqnarray}
The average $\langle q_{iz} \rangle$ is
\begin{eqnarray}
\langle q_{iz} \rangle &=&  \frac{p_0/n-m_{gT}}{\xi_{\rm max}}.
\label{72}
\end{eqnarray}
With $p_0=10$ GeV/c and $m_{gT}=0.6$ GeV, we find for $n=6$,
\begin{eqnarray}
 q_{iz} \sim 0.6 {\rm~~GeV/c},
\label{78}
\end{eqnarray}
and for 
for $n=2.4$, we find
\begin{eqnarray}
 q_{iz} \sim 1.4 {\rm~~GeV/c}.
\label{79}
\end{eqnarray}
These estimates indicate that the average longitudinal momentum kick
acquired by a gluon scatterer is slightly smaller than that acquired
by a fermion scatterer. They are approximately inversely proportional
to the number of scatterers in a coherent collision.

\section{Signatures of Bose-Einstein Interference and Comparison with Experimental data}

The results in the above sections provide information on the signatures
for the occurrence of the Bose-Einstein interference in the coherent
collisions of a jet with medium partons:

\begin{enumerate}

\item
The Bose-Einstein interference is a quantum many-body effect.  It
occurs only in the multiple collisions of the fast jet with two or
more scatterers.  Therefore there is a threshold corresponding to the
requirement of two or more scatterers in the multiple collisions,
$n\ge 2$.

\item
Each scatterer has a transverse momentum distribution 
of the type $1/|{\bb q}_T|^4$, which peaks at small values of $|{\bb q}_T|$.

\item
Each scatterer acquires a longitudinal momentum kick 
that is of order $1/2n$ fraction of the incident jet momentum 
along the incident jet direction .

\item
As a consequence, the final effect is the occurrence of collective
recoils of the scatterers along the jet direction.

\end{enumerate}

To inquire whether Bose-Einstein interference may correspond to any
observable physical phenomenon, it is necessary to identify the
scatterers to separate them from the incident jet in a measurement.
Such a separation is indeed possible in $\Delta \phi$-$\Delta\eta$
angular correlation measurements of produced pairs with a high-$p_T$
trigger \cite{Ada05}-\cite{ ALI12}.  Particles in the ``ridge" part
of the correlations with $|\Delta \eta| > 0.6$ at RHIC and 
$|\Delta \eta| > 1.0$ at LHC
with $\Delta \phi \sim
0$ can be identified as belonging to the medium partons because
it was observed at RHIC that 

\begin{enumerate}

\item
The yield of these ridge particles increases approximately linearly with the number of participants \cite{Put07}.

\item
The yield of these ridge particles is nearly independent of (i) the
flavor content, (ii) the meson/hyperon character, and (iii) the
transverse momentum $p_T$ (above 4 GeV) of the jet trigger
\cite{Put07,Bie07,Bie07a}.

\item
The ridge particles have a temperature (inverse slope) that is similar
(but slightly higher) than that of the inclusive bulk particles, but
lower than the temperature of the near-side jet fragments \cite{Put07}

\item
The baryon/meson ratio of these ridge particles is similar to those of
the bulk hadrons and is quite different from those in the jet
fragments \cite{Lee09}.

\end{enumerate}

With the scatterers as ridge particles that can be separated from the
incident high-$p_T$ jet, the occurrence of the Bose-Einstein
interference will be signaled by Item (4) of the collective recoils of the
scatterers (the ridge particles) along the jet direction.  Such
collective recoils will lead to the $\Delta \phi \sim 0$ correlation
of the ridge particles with the high-$p_T$ trigger, as has been
observed in angular correlations of produced hadrons in AuAu
collisions at RHIC \cite{Ada05}-\cite{Jia08qm},
and in pp and PbPb collisions at LHC \cite{CMS10,CMS11,CMS12,ATL12,ALI12}.
The collective
recoils of the kicked medium partons have been encoded into the
longitudinal momentum kick $\langle q_{iz} \rangle$ of the momentum
kick model that yields the observed $\Delta \phi$, $\Delta \eta$, and
$p_T$ dependencies of the angular correlations in AuAu collisions at RHIC
\cite{Won07,Won08ch,Won08,Won08a,Won09,Won09a,Won09b},
and pp collisions at LHC \cite{Won11}.

It is of interest to examine Item (3) of the signature of the
Bose-Einstein interference with regarding to the relationship between
the (average) magnitude of the longitudinal momentum kick, $\langle
q_{iz} \rangle$, and the (average) number of scatterers, $\langle n
\rangle$, when such a collective momentum kick occurs.  For the most
central AuAu collisions at $\sqrt{s_{NN}}=200$ GeV at RHIC, we
previously found that $\langle f_R \rangle \langle n \rangle \sim 3.8$
where $n$ is the number of kicked medium partons and $\langle f_R
\rangle$ is the average attenuation factor for the kicked partons to
emerge from the collision zone \cite{Won08}.  The value of $\langle
f_R \rangle$ is not determined but a similar attenuation factor
$\langle f_J \rangle$ for jet fragments is of order $0.63$ \cite{Won08}.  We can therefore estimate that for the most central
AuAu collisions at $\sqrt{s_{NN}}=200$ GeV at RHIC, $ \langle n
\rangle \sim 6$.  For an incident jet of $p_z$$\sim$10 GeV/c
\cite{Ada08},
the estimates of Eq.\ (\ref{38})  and (72) give
\begin{eqnarray}
q_{iz}\sim 
\begin{cases}
{  0.83  {\rm ~GeV/c} & ~~{\rm for~a~quark~scatterer}, \cr 
  0.63  {\rm ~GeV/c} &~~{\rm for~a~gluon~scatterer.} \cr
}
\end{cases}
\label{80}
\end{eqnarray}
These estimates of the momentum
kick are of the same order as   the value of $\langle q_{iz}\rangle \sim
1$ GeV/c estimated in \cite{Won08} and 0.8 GeV/c in \cite{Won09},
obtained in the momentum kick model analysis.

In another momentum kick model analysis for the highest multiplicity
$pp$ collisions at $\sqrt{s_{NN}}=7$ TeV at the LHC, we previously
found that $\langle f_R \rangle \langle n \rangle \sim 1.5$
\cite{Won11}. 
The value of ridge particle attenuation  factor $\langle
f_R \rangle$ is not determined but we can use again the similar attenuation factor
for jet fragments, $\langle f_J \rangle$ of order $0.63$ \cite{Won08},
to estimate $\langle n \rangle \sim 2.4$.  For an incident jet of 
10 GeV/c, the
estimates of Eqs.\ (\ref{38}) and (\ref{72}) give
 the average scatterer longitudinal recoil momentum 
 as
\begin{eqnarray}
q_{iz} \sim
\begin{cases}
{  2.1  {\rm ~GeV/c} & ~~{\rm for~a~quark~scatterer}, \cr 
  1.4  {\rm ~GeV/c} &~~{\rm for~a~gluon~scatterer.}  \cr
}
\end{cases}
\end{eqnarray}
 These estimates of the momentum
kick are slightly lower than but are of the same order as  
of the same order as the value of $\langle q_{iz}\rangle \sim
2$ GeV/c inferred from experimental data   in the momentum kick model analysis \cite{Won11}.  The
experimental data give a longitudinal momentum transfer that is
approximately inverse proportional to the number of scatterers.

\begin{figure} [h]
\includegraphics[scale=0.35]{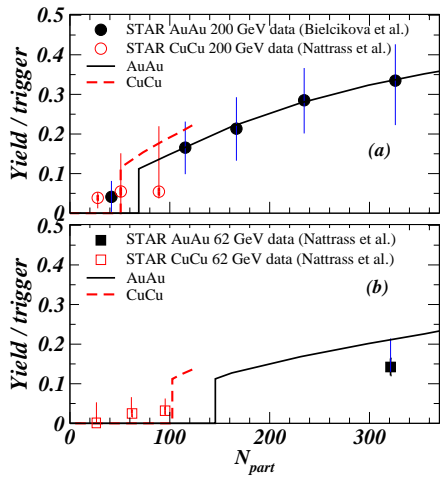}
\caption{(Color online) The ridge yield per high-$p_T$ trigger as a
  function of the participant number $N_{\rm part}$ for
  nucleus-nucleus collisions at $\sqrt{s_{NN}}$=200 and 62 GeV.  The
  solid circular points (for AuAu) and the square points (for CuCu)
  are from the STAR Collaboration \cite{Bie07,Nat08}.  The curves are
  the momentum kick model results of \cite{Won08a} modified to include
  the Bose-Einstein interference threshold effect of $n \ge 2$. }
\end{figure}

With regard to Item (1) for the signature for the occurrence of the
Bose-Einstein interference, the presence of a threshold implies a
sudden increase of the ridge yield as a function of the number of kicked medium scatterers, $n$, which increases with the increasing centrality, as
represented either by an increasing number of participants or by increasing multiplicities.  
The slope of the yield as a function of centrality will change from zero below threshold to infinite over a small range of centralities at threshold.  We expect that such a  sudden threshold will be smoothed out by  
the fluctuations of the participant numbers 
and  multiplicities with respect to the number of kicked medium partons.   However the change of the slope of the ridge yield as a function of centralities will remain.  Hence,  the BE threshold effect will show up as a change of the slope of the ridge yield as a function of centralities, measured by the number of participants or by multiplicities.  Equivalently, the ridge yield as a function of centralities will appear to have a kink near the threshold region of centrality.  A search for a change of the slope or a kink in the ridge yield as a function of centralities will allow us to find the BE interference threshold.

We show in Fig.\ 5 the
experimental ridge yield per high-$p_T$ trigger as a function of
$N_{\rm part}$ for AuAu and CuCu collisions at $\sqrt{s}=200$ and $62$
GeV at RHIC \cite{Bie07,Nat08}.  We also show in Fig.\ 5 the
theoretical yields obtained in the momentum kick model \cite{Won08},
where the ridge yield at the most central collision at $N_{\rm
  part}=320$ was calibrated as $n=6$ \cite{Won08}.  With such a
calibration, the threshold values in $N_{\rm part}$ at which $n=2$ can
be located and listed in Table I, where the $30\%$ theoretical errors
arise from the errors in measuring the ridge yield at the most central
collision at $N_{\rm part}=320$.  Theoretical ridge yields from the
momentum kick analysis in Fig.\ 10 of Ref.\ \cite{Won08}, modified to
include the Bose-Einstein interference threshold effect of $n \ge 2$,
are shown as the solid curves for AuAu collisions, and as dashed
curves for CuCu collisions in Fig.\ 5.  The theoretical thresholds in
Fig.\ 5 will be smoothed out by the fluctuations of the number of
scatterers as a function of $N_{\rm part}$
and by uncertainties in the estimates of
the number of scatterers.  Although the experimental
data appear to be consistent with theory, the large error bars and the scarcity of the number of
data points in the threshold regions preclude a definitive conclusion.

\begin{figure} [h]
\includegraphics[scale=0.45]{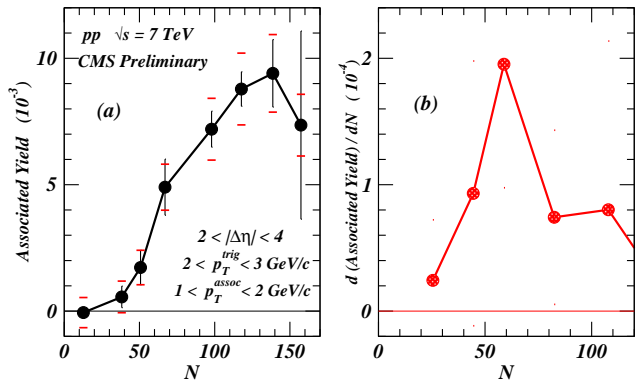}
\caption{ (Color online) (a) Preliminary CMS Collaboration data of the ridge yield per trigger in the region of $2<|\Delta \eta| < 4$
for pp collisions at 7 TeV, for $2<p_T^{\rm trig}<3$ GeV, 
  $1<p_T^{\rm assoc}<2$ GeV, as a function of multiplicities $N$ \cite{CMS12b}.  (b) The slope of the ridge yield,
 $d({\rm Associated~Yield})/dN$ as a function of $N$.
}
\end{figure}

In Fig.\ 6(a), we show the CMS preliminary data on the near-side ridge yield per trigger as a function of centralities, as measured by multiplicities $N$,
for $pp$ collisions at $\sqrt{s}$=7 TeV  \cite{CMS12b}.
To search for the BE threshold that may show up as  a change of slope or a kink of  the ridge yield as a function of multiplicities, we show  the slope of the associated ridge yield, $d({\rm Associated~Yield})/dN$ as a function of $N$
in Fig.\ 6(b).   The lines in Fig.\ 6(a) and 6(b)  join the data points to guide the eyes.
  The CMS preliminary data in Fig.\ 6(a) and 6(b)  indicate a sharp change of the slope of the ridge yield in the region around 
 $N\sim 50-70$,  which may suggest a threshold for the ridge yield at around $N=50-70$.  

We show CMS preliminary data \cite{CMS12b} on the near-side ridge yield per trigger for PbPb collision at $\sqrt{s_{NN}}$=2.76 GeV/c   as a function of the number of participants $N_{\rm part}$ in Fig.\ 7(a), and the corresponding $d({\rm Associated~Yield})/dN_{\rm part}$  in Fig.\ 7(b).  In Fig.\ 7, we also included the minimum-biased pp data point as an open circle for $N_{\rm part}=2$  (for very peripheral PbPb collisions), at which the ridge yield is zero \cite{CMS10,CMS12}.
The integrated near-side associated yield for $4 < p_T^{\rm trig}<6$ GeV/c and $2 < p_T^{\rm assoc}<4$ GeV in the ridge region of $2<|\Delta \eta|<4$ appear to have a kink in the ridge yield as a function of $N_{\rm part}$ 
at $\langle N_{\rm part} \rangle$$\sim$30 in Fig.\ 7(a).
The slope increases sharply starting at $N_{\rm part}\sim 20$ in Fig.\ 7(b) and reaching a large value at $N_{\rm part}\sim 40$.  Such a behavior may suggest  the presence of a ridge yield threshold.
The location of a possible threshold at $\langle N_{\rm part} \rangle$$\sim$30 for PbPb collisions appear to be  qualitatively consistent with the decrease of $\langle N_{\rm part} \rangle$ as a function of increasing collision energies, as indicated in Table I.  It will be of interest to examine 
in future work whether the  ridge
 thresholds as suggested by the CMS   pp and PbPb 
data  correspond quantitatively to the location of $n=$2  for the onset of the BE interference.
\begin{figure} [h]
\includegraphics[scale=0.45]{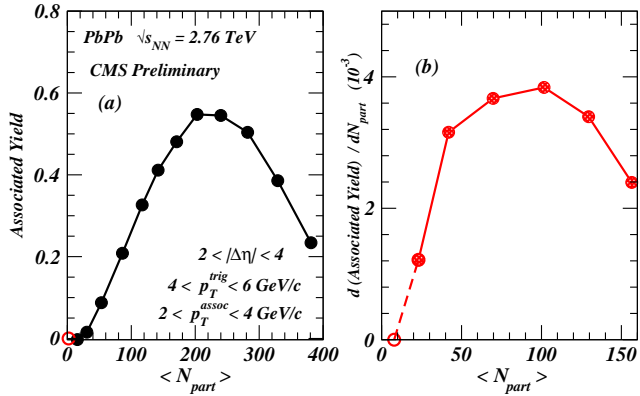}
\caption{ Preliminary CMS data of the ridge yield per trigger in the ridge region of $2<|\Delta \eta| < 4$
for PbPb collisions at $\sqrt{s_{NN}}=$2.76 TeV, for $4<p_T^{\rm trig}<6$ GeV, 
  $2<p_T^{\rm assoc}<4$ GeV, as a function of the number of participants $N_{\rm part}$. \cite{CMS12b}.}
\end{figure}

Threshold effects for the ridge
yield (2D Gaussian yield) as a function of $N_{\rm part}$ have been
observed in another angular correlation measurements with a low-$p_T$
trigger from the STAR Collaboration \cite{Dau08,Ray08a,Ket09,STA11} as shown in Fig.\ 8.  We
note previously that a fast jet parton possesses low-$p_T$ jet
fragments and a minimum-$p_T$-biased low-$p_T$ trigger can also
indicate the passage of a fast parent jet \cite{Won11}.  As a
consequence, ridge particles will also be associated with a low-$p_T$
trigger.   The change of the slope of the
amplitude of the 2D Gaussian distribution (the ridge yield)  shown in Fig.\ 8 indicates the
presence of a threshold for the ridge yield as a function of
centrality.

\begin{table}[h]
 \caption {  Comparison of the locations of the
   theoretical threshold, at which the (average) number of scatterer
   $n$ is equal to 2, with the observed experimental threshold
   \cite{Dau08,Ray08a,Ket09,STA11} for the sudden increase of the 
2D Gaussian distribution 
   amplitude and width  (ridge component)
   in AuAu collisions at $\sqrt{s_{NN}}=200$ and 62 GeV in Fig.\ 8.}
\vspace*{0.2cm} 
\hspace*{0.0cm}
\begin{tabular}{|c|c|c|c|}
\hline    
Collision & $\sqrt{s_{NN}}$     &  Theoretical   & Experimental  \\
System  &  (GeV)    &  Threshold $N_{\rm part}$    &  Threshold $N_{\rm part}$ \\ \hline 
AuAu &    200  &  69$\pm$21        &      58-86            \\ 
AuAu &     62  &  146$\pm$45        &     85-122           \\ \hline 
CuCu &    200  &  51$\pm$15        &                 \\ 
CuCu &     62  &  103$\pm$31        &               \\ \hline 
\end{tabular}
\end{table}

\begin{figure} [h]
\includegraphics[scale=0.55]{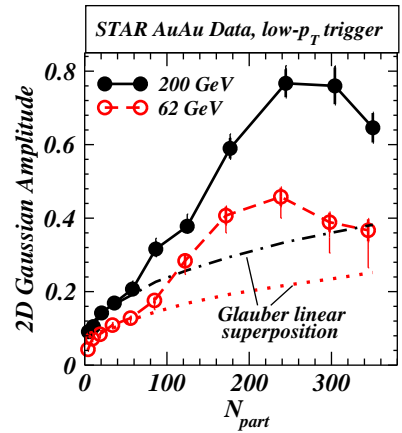}
\caption{ (Color online) STAR Collaboration data of  the 
2D Gaussian distribution
  amplitude (ridge
  yield)   for AuAu collisions at
  $\sqrt{s_{NN}}$=200 GeV (solid circles) and 62 GeV (open circles),
  with a low-$p_T$ trigger \cite{Dau08,Ray08a,Ket09,STA11}.  The solid
  and dashed curves are lines joining the data points.  The dashed-dot
  and dotted curves represent Glauber linear superposition (GLS)
  estimates.}
\end{figure}

For AuAu at 200 GeV/c, the experimental ridge yield threshold occurs
at $N_{\rm part}= 58-86$ (Fig.\ 8), which can be compared in Table I
with the theoretical ridge yield threshold of $N_{\rm part}= 69\pm 21$
as estimated in the momentum kick model for $n=2$.  For AuAu at 62
GeV/c, the experimental ridge yield threshold occurs at $N_{\rm part}=
85-122 $ (Fig.\ 8), which can be compared with the theoretical
threshold of $N_{\rm part}= 146\pm 45$ as estimated in the momentum
kick model for $n=2$ (Table I).  These comparisons indicate that
experimental data with low-$p_T$ trigger are consistent with the
presence of ridge thresholds as a function of the number of the medium
scatterers located at $n\sim 2$, as in the threshold effect in
Bose-Einstein interference.

\section{Conclusions and Discussions}

Conventional investigations 
\cite{Gyu94,Bai96,Wie00,Gyu01,Djo04,Ova11,Meh11,Arm12,Gla59}
on scattering of a fast particle with medium scatterers use the potential model in which the scatterers 
are represented by static potentials and the longitudinal recoils of the scatterers are considered as dependent variables that are appendages to the deflected  motion of the incident particle.  
A general treatment of
 coherent collisions
necessitates the use of
the 
longitudinal recoils of the scatterers
as
independent dynamical variables which are 
however
not allowed
in the potential model. 
This leads us to forgo the potential model and to turn to the use of
Feynman amplitudes for the general treatment of coherent collisions.

In the Feynman amplitude approach for the coherent collision
of a fast particle on $n$ scatterers, there are $n!$ different orderings in the sequence of collisions along 
the fast particle trajectory
at which various virtual bosons are exchanged.
 By Bose-Einstein
symmetry, the total Feynman amplitude is the sum of the $n!$
amplitudes for all possible interchanges of the virtual
bosons.  The summation of these $n!$ Feynman amplitudes and the
accompanying interference constitute the Bose-Einstein interference in
the passage of the fast particle in the dense medium.

Our interest in examining this problem has been stimulated by the phenomenological successes of the momentum kick
model in the analysis of the angular correlations of hadrons produced
in high-energy heavy-ion collisions
\cite{Won07,Won08ch,Won08,Won08a,Won09,Won09a,Won09b,Won11}. We seek a
theoretical foundation for the origin of the longitudinal momentum
kick along the jet direction postulated in the model.  We explore whether such a longitudinal
momentum kick may originate from a quantum many-body effect arising
from the Bose-Einstein interference in the passage of a jet in a dense
medium.  We take note of previous results on the Bose-Einstein
interference in the emission of real photons and gluons in high-energy
interactions and in the sum of the ladder and cross-ladder loop
diagrams in the collision of two particles
\cite{Che69,Che87,Fen96,Fen97,Lam96a,Lam97,Lam97a,Lam97b}.

We find similarly that in the coherent collisions of an energetic
fermion with $n$ fermion scatterers at high energies in the Abelian
theory, the symmetrization of the Feynman scattering amplitudes with respect to
the interchange of the virtual bosons leads to the Bose-Einstein
interference, resulting in a sharp distributions at $p\cdot q_i\sim
0$.  Such coherent collisions are in fact a single collision, tying
the incident fermion with the $n$ fermion scatterers as a single unit.
There are then $3(n+1)$ degrees of freedom, subject to the constraints
of the conservation of energy and momentum.  As a consequence, all
$3n$ degrees of freedom of the scatterers can be independently varied.
The probability distribution in these $3n$ momentum degrees of freedom depends
on the
Feynman amplitudes and the phase-space factors.  The Bose-Einstein symmetry constraints of $p\cdot
q_i\sim 0$ limit the transverse momentum transfers of the scatterers
to small values of $q_{iT}$.  The longitudinal momenta of the
scatterers get their share of longitudinal momenta from the jet,
resulting in the collective recoils of the scatterers along the jet
direction.

For the coherent collision of an energetic parton with parton
scatterers in non-Abelian cases, we find that the complete
Bose-Einstein symmetry in the exchange of virtual gluons consists not
only of space-time exchange symmetry but also color index exchange
symmetry.  Nevertheless, there is always a space-time symmetric and
color-index symmetric component of the Feynman amplitude that behaves in
the same way as the Feynman amplitude in the Abelian case, in addition
to the occurrence of space-time antisymmetric and color-index
antisymmetric components.  For the space-time symmetric and
color-index symmetric component, the recoiling partons behave in the
same way as in collisions in the Abelian case.  There is thus a finite
probability for the parton scatterers to emerge collectively along the
incident trigger jet direction, each with a significant fraction of
the longitudinal momentum of the incident jet.  The collective recoils
will lead to the $\Delta \phi \sim 0$ correlation of the ridge
particles with the high-$p_T$ trigger.  Such a signature of the
Bose-Einstein interference may have been observed in the $\Delta
\phi\sim 0$ correlation of the ridge particles with a high-$p_T$
trigger in the angular correlation measurements of produced hadron
pairs in AuAu collisions at RHIC \cite{Ada05}-\cite{Jia08qm}
and in pp and  PbPb collisions at LHC \cite{CMS10,CMS11,CMS12,ATL12,ALI12}.
 The
centrality dependence of the ridge yields in 
pp and PbPb collisions at LHC with high $p_T$ trigger
\cite{CMS12,CMS12b}
and  in AuAu collisions at RHIC 
with a low-$p_T$ trigger
\cite{Dau08,Ray08a,Ket09,STA11} gives hints of the presence
of a ridge threshold, as expected in the quantum many-body
effect of Bose-Einstein interference.

Our focus at present is on the recoils of the scatterers.  
We have been examining the collision between the incident particle and the medium scatterers without radiation.
We find that the scatterer recoils ranges from quasi-elastic to substantial fraction of the incident incident particle momentum.
For a given multiple collision processes involving the recoils of the scatterers,
 the radiative processes will involve addition external legs
in the Feynman diagram.
 They 
are high order in $\alpha_s$.     Therefore 
as far as the cross sections are concerned,
they generally occur with less probabilities.  
However, as far as jet energy loss is concerned, these high-order radiative processes can be important in  
certain kinematic regions.
For example, if one restricts oneself to quasi-elastic processes with little scatterer longitudinal momentum recoils, then radiation can take up more of the jet
longitudinal momentum, even though they occur with a lower probability than quasi-elastic scattering.  One therefore envisages that in the investigation of  the loss of the incident longitudinal momentum,  radiative energy loss is  important in the region of small scatterer longitudinal recoils, but 
the relative importance of the  radiative energy loss will diminish  as one moves toward the regions of greater scatterer longitudinal recoils.

If one restricts oneself to potential models, then the longitudinal momentum and energy losses due to scatterer recoils will be small and radiative energy loss becomes more important than quasi-elastic collision energy loss.
However, for reasons we examine in Section II and discussed earlier in this section, the general treatment of
 coherent collisions
necessitates the use of
the 
longitudinal recoils of the scatterers
as
independent dynamical variables, which are 
however
not allowed
in the potential model.   It is necessary to use the Feynman amplitude approach to explore
  the entire domain of longitudinal scatterers recoils.
 We hope to carry out an analysis of radiative processes
in conjunction with scatterer recoils  within the Feynman amplitude approach in the future.

The present work is based on the high-energy limit which allows great
simplifications of the algebraic structures of Feynman
amplitudes. These simplified structures bring into clear focus the
mechanism of the Bose-Einstein interference that changes the nature of
the distribution function.  With the mechanism of the Bose-Einstein
interference well understood, it may be beneficial in future work to
evaluate the Feynman amplitudes without resorting to many of the
drastic assumptions and simplifications -- in order to make
quantitative comparison of theoretical predictions with results of the
longitudinal momentum kick quantities extracted from experimental
data.  This is particularly important in regions of large longitudinal
momentum transfers for which the high-energy approximation of having
$p'$ not greatly different from $p$ may not hold.

Coherence or decoherence in high energy processes occurs when the scattering amplitude consists of many contributions and these contributing amplitudes interfere.  Because of its general nature, they come  in many different processes in different forms.  

The coherence results we present here 
have been  obtained for  a single
jet and interacting particles  as plane waves. 
High-energy collisions has a preference for longitudinal
motion over transverse motion.  Therefor, for a single jet, as in the case of our main interest, the interference of Feynman diagrams has relevance
only with regard to longitudinal coherence length along the
longitudinal direction. 
For such a configuration, the uncertainties of the vertices of boson emission along the longitudinal direction leads to the coherence in the collision process.  Uncertainties in the transverse direction come from the transverse positions of the scatterers.  Such uncertainties can be included by treating the scatterers in terms of  wave packets \cite{Pes95}, form factors  \cite{Gla70}, or  scatterer wave function  \cite{Gla59}.  In the special case of a purely elastic scattering of 
an incident fast particle with
medium particles (in a nucleus), there can be constructive interference of the scattering amplitudes from different particles, leading to a sharp diffractive elastic peak at the forward angle, as described by Glauber \cite{Gla70}.

 However, there are circumstances in which
interesting transverse uncertainties arises, and the amplitudes from different transverse sources interfere.
An interesting case of transverse interference  can be found in connection with  the 
coherent emission of a gluon
 from  a system of a quark jet and  an antiquark jet \cite{Meh11,Arm12}.
 This case 
involves the propagation of two jets  with an opening transverse angle $\theta_{q\bar q}$ and
differs from our present case with the propagation of a single jet.   The amplitude for gluon emission from the quark $q$  interfere with the amplitude for gluon emission from the antiquark $\bar q$.  As a consequence,  in the absence of medium, the coherent interference in the transverse direction lead to the angular ordering of the emitted gluon in such a way that the emitted gluon lies within the opening angles of the $q$-$\bar q$ pair,  $\theta_{q\bar q}$.    In the presence of a medium,  as shown in Ref.\ \cite{Meh11,Arm12}, the quark  and the antiquark  are subject to  multiple scatterings along their trajectories that change their  momenta, their distributions, and their propagating phases.  These changes of the quark and antiquark sources of the gluon due to the medium will diminish the strength of the transverse coherent interference 
of the emitted gluon that occurs in the absence of the medium,
leading to a gradual decoherence of the gluon emission as a function of increasing medium density \cite{Meh11,Arm12}.   Recent experimental findings on reconstructed jets in nuclear
collisions at the LHC \cite{ATL11b,CMS11b} suggest that such medium-induced partial decoherence may be an important effect.

Conventional model analysis of collision processes 
\cite{Gyu94,Bai96,Wie00,Gyu01,Djo04,Ova11,Meh11,Arm12,Gla59}
assumes the ladder-type diagrams such as Fig. 1(a) and Fig.\ 2(a), while the cross Feynman diagrams such as Fig.\ 1(b) and Figs.\ 2(b)-2(f) have been ignored.  As the derivations in Section III demonstrate, 
the contributions from different cross diagrams destructively interfere with the ladder Feynman diagram and  there is a high degrees of cancellation.
In general, the neglect of these destructive interference by including only the ladder Feynman diagram cannot  be justified from a mathematical view point.
Only in the special and restrictive case of quasi-elastic scattering with  the fast particle assumed to be nearly on the mass shell and with the additional assumption of $q_{i0}=0$ is it justified to include 
only the ladder diagrams, at the expenses of 
precluding the exploration into other regions  
of scatterer recoils.

Many models have been proposed to explain the ridge phenomenon
\cite{Won07,Won08ch,Won08,Won08a,Won09,Won09a,Won09b,Won11,
  Shu07,Vol05,Chi08,Hwa03,Chi05,Hwa07,Pan07,Dum08,Gav08,Gav08a,
  Arm04,Rom07,Maj07,Dum07,Miz08,Jia08,Jia08a,Ham10,Dum10,Wer11,Hwa11,Chi10,
  Tra10,Tra11a,Tra11b,Tra11c,Sch11,Pet11,Arb11,Aza11,Hov11,Bau11,Che10,Dre10,Lev11}.
They include the collision of jets with medium partons
\cite{Won07,Won08ch,Won08,Won08a,Won09,Won09a,Won09b,Won11,Hwa11,Chi10},
flows and hydrodynamics with initial state fluctuations
\cite{Shu07,Vol05,Gav08,Gav08a,Ham10,Wer11,Sch11}, color-glass
condensate \cite{Dum07,Dum08,Gav08,Gav08a,Dum10}, modeling pQCD
\cite{Tra10,Tra11a,Tra11b,Tra11c}, parton cascade \cite{Pet11}, gluon
bremsstrahlung in string formation \cite{Arb11}, strong-coupling
AdS/CFT \cite{Hov11}, quantum entanglement \cite{Che10}, and BFKL
evolution and beyond \cite{Lev11}.  There are however two difficulties
associated with these phenomenological models.  Almost all models deal
with fragmented parts of the data and all models contain implicit and
explicit assumptions.

In the presence of a large number of models and the above
difficulties, progress can proceed in three fronts.  First, the models
need to cover an extensive set of experimental differential data over
large phase spaces, centralities, and energies, from many different
collaborations.  Secondly, the assumptions of the models need further
theoretical and observational investigations from fundamental
viewpoints.  Finally, experimental tests need to be proposed to
distinguish different models.

With regard to the first of these three tasks, the momentum kick model
gives reasonable descriptions for an extensive set of differential
data of the ridge yield, over an extended range of transverse momenta,
azimuthal angles, pseudorapidity angles, centralities, and collision
energies, from the STAR Collaboration
\cite{Ada05,Ada06,Put07,Bie07,Wan07}, the PHENIX Collaboration
\cite{Ada08,Mcc08,Che08,Jia08qm}, the PHOBOS Collaboration
\cite{Wen08}, and the CMS Collaboration \cite{CMS10}.  In these
momentum kick model analyses, a collective longitudinal momentum kick
on the medium scatterers along the jet direction is a central
ingredient leading to the successful descriptions of the large set of
experimental data.  Additional analyses will be continued to extend
the momentum kick model to include the effects of collective flows and
to cover a larger set of new data as they become available.

With regard to the second of the three tasks, the momentum kick
model contains the basic assumption that the $\Delta \eta$ ridge
arises mainly from the initial rapidity distribution of partons prior
to the jet collision.  Such a basic assumption has been examined from
the viewpoint of the Wigner function  of produced particles in a
fundamental quantum theory of particle production \cite{Won09a}.
Another basic assumption concerning the longitudinal nature of the
momentum kick is now being examined in the present manuscript.

With regard to third of the three tasks, the present analysis for jets
interacting with medium partons reveals that the longitudinal momentum
kick is a quantum many-body effect that contains a threshold,
requiring the multiple collisions of the jet with at least two
partons. This may be in agreement with the onset of the ridge yield
as a function of the centrality, observed by the CMS Collaboration for pp and PbPb collisions at LHC \cite{CMS12b},  and  
by the 
 STAR Collaboration at RHIC \cite{Ada06,Tra08,Dau08,Ray08a,Ket09,STA11}, as discussed in Section X.
Furthermore, in the momentum kick model with longitudinal momentum
kicks, the kicked medium partons from back-to-back jets possess a
$(p_{1T}, p_{2T})$ correlation for both the near side and the away
side, as observed by the STAR Collaboration \cite{Tra08,Won11}.

It would be of interest to see how other proposed models fare with the
three tasks at hand.  It will also be of interest to see whether they
contain the features of a  rapid rise  of the ridge yield as a
function of the centrality, and the $(p_{1T}, p_{2T})$ correlation for
both the near side and the away side.  The search for tests to
distinguish different models will be an on-going research activity.

Whatever the theoretical descriptions, jets of order 10 GeV and
below are known to be present in high-energy collisions
\cite{Wan92,Ada08,Mcc08,Che08,Jia08qm,Tan09}.  These jets collide
with medium partons.  They are not hydrodynamical flows, but they
contribute to the azimuthal anisotropy and the azimuthal Fourier
coefficients \cite{Won79,Oll92,Vol96}, as well as to two-particle
correlations.  They must be taken into account or be subtracted from
experimental data in theoretical models that do not include the jet effects explicitly.  

In summary, Bose-Einstein interference in the passage of a jet in a
dense medium is a quantum many-body effect that occurs quite generally
in a coherent multiple collision process, with the threshold of more
than two scatterers.  The manifestation of the Bose-Einstein
interference effect as collective recoils of the scatterers along the
jet direction may have been experimentally observed in the $\Delta
\phi$$\sim$0 correlation of hadrons associated with a high-$p_T$
trigger in high-energy AuAu collisions at RHIC, and 
pp and PbPb collisions at LHC.   The experimental observation of ridge thresholds as a
function of centrality may lend support for the occurrence of
the Bose-Einstein interference thresholds in the passage of a jet in a
dense medium.

\vspace*{0.3cm}

\centerline{\bf Acknowledgment}

\vspace*{0.3cm} The author wishes to thank Profs.\ C. S. Lam,
H. W. Crater, Jin-Hee Yoon, Vince Cianciolo, A. V. Koshelkin, and Wei Li,
for helpful discussions.  This research was supported
in part by the Division of Nuclear Physics, U.S. Department of Energy.


\begin{thebibliography}{99}



\bibitem{Ada05}
J. Adams $et~al.$ for the STAR Collaboration, 
Phys. Rev. Lett. {\bf 95}, 152301 (2005).

\bibitem{Ada06}
J. Adams $et~al.$ (STAR Collaboration), 
Phys. Rev. C {\bf 73}, 064907 (2006).

\bibitem{Put07} J. Putschke (STAR Collaboration), 
J. Phys. {\bf G34}, S679 (2007).

\bibitem{Bie07} J. Bielcikova (STAR Collaboration), 
J. Phys. {\bf G34}, S929 (2007).

\bibitem{Wan07} 
F. Wang (STAR Collaboration), Invited talk at the XIth
International Workshop on Correlation and Fluctuation in Multiparticle
Production, Hangzhou, China, November 2007, [arXiv:0707.0815].

\bibitem{Bie07a}
 J. Bielcikova (STAR Collaboration), Phys.G34:S929-930,2007;
 J. Bielcikova for the STAR Collaboration, Talk
presented at 23rd Winter Workshop on Nuclear Dynamics, Big Sky,
Montana, USA, February 11-18, 2007, [arXiv:0707.3100];
J. Bielcikova for the STAR Collaboration, Talk presented at XLIII
Rencontres de Moriond, QCD and High Energy Interactions, La Thuile,
March 8-15, 2008, [arXiv:0806.2261].

\bibitem{Abe07} B. Abelev (STAR Collaboration), Talk presented at 23rd
Winter Workshop on Nuclear Dynamics, Big Sky, Montana, USA, February
11-18, 2007, [arXiv:0705.3371].

\bibitem{Mol07}
L. Molnar (STAR Collaboration), J. Phys. G {\bf 34}, S593 (2007).

\bibitem{Lon07} R. S. Longacre (STAR Collaboration),
  Int. J. Mod. Phys. E{\bf 16}, 2149 (2007).


\bibitem{Nat08} C. Nattrass (STAR Collaboration), 
J. Phys. G {\bf 35}, 104110  (2008).  

\bibitem{Fen08}
A. Feng, (STAR Collaboration), 
J. Phys. G {\bf 35}, 104082  (2008).

\bibitem{Net08} P. K. Netrakanti (STAR Collaboration)
J. Phys. G {\bf 35}, 104010  (2008).

\bibitem{Bar08}
O. Barannikova (STAR Collaboration), 
J. Phys. G {\bf 35}, 104086  (2008).

\bibitem{Dau08}
M. Daugherity, 
(STAR Collaboration), 
J. Phys. G {\bf 35}, 104090  (2008).

\bibitem{Ray08a}
D. Ray, in Talk presented at
Tamura Symposium on
Heavy Ion Physics,
the University of Texas at Austin,  November 20–22, 2008,
http://www.ph.utexas.edu/~molly/tamura/.


\bibitem{Ket09}
D. Kettler, (STAR Collaboration), 
Euro. Phys. Jour. C, {\bf 62}, 175  (2009).

\bibitem{STA11}
G. Agakishiev $et~al.$, (STAR Collaboration),
arxiv:1109.4380 (2011).


\bibitem{Tra08}
 T. A. Trainor  Phys. Rev {\bf C78}, 064908 (2008);
T. A. Trainor and D. T. Kettler, Phys. Rev. D 74, 034012 (2006);
T. A. Trainor, Phys. Rev. C 80, 044901 (2009);T. A. Trainor,
J. Phys. G 37, 085004 (2010).

\bibitem{Lee09}
M. van Leeuwen, (STAR Collaboration), 
Eur. Phys. J. C {\bf 61}, 569 (2009).


\bibitem{Ada08}
A. Adare, $et~al.$ (PHENIX Collaboration), 
Phys. Rev. C {\bf 78}, 014901 (2008).

\bibitem{Mcc08}
M. P. McCumber  (PHENIX Collaboration), 
J. Phys. G {\bf 35}, 104081 (2008).

\bibitem{Che08} Chin-Hao Chen (PHENIX Collaboration), 
``Studying the Medium Response by Two Particle Correlations", Hard Probes 2008
  Intern. Conf. on Hard Probes of High Energy Nuclear Collisions, A
  Toxa, Galicia, Spain, June 8-14, 2008.


\bibitem{Jia08qm}
Jiangyong Jia, (PHENIX Collaboration), 
J. Phys. G {\bf 35}, 104033 (2008).

\bibitem{Tan09}
M.J. Tannenbaum, 
Eur. Phys. J. C {\bf 61}, 747 (2009).


\bibitem{Wen08} E. Wenger (PHOBOS Collaboration), J. Phys. G {\bf 35},
  104080 (2008).

\bibitem{CMS10}
CMS Collaboration, JHEP 1009, 091 (2010),[arxiv:1009.4122].

\bibitem{CMS11}
CMS Collaboration, arXiv:1105.2438 (2011).

\bibitem{CMS12}
CMS Collaboration, arXiv:1201.3158 (2012).

\bibitem{ATL12}
ATLAS Collaboration, arXiv:1203.3087 (2012).

\bibitem{ALI12}
ATLAS Collaboration,
Phys. Lett. {\bf B708},   249 (2012). 

\bibitem{Won07}
C. Y. Wong, Phys. Rev. C {\bf 76}, 054908  (2007).

\bibitem{Won08ch} C. Y. Wong, Chin. Phys. Lett. {\bf 25}, 3936 (2008).

\bibitem{Won08} C. Y. Wong, J. Phys. G {\bf 35}, 104085 (2008).

\bibitem{Won08a}
C. Y. Wong, Phys. Rev. C {\bf 78}, 064905 (2008).

\bibitem{Won09}
C. Y. Wong, Phys. Rev. C {\bf 80}, 034908 (2009). 

\bibitem{Won09a}
C. Y. Wong, Phys. Rev. C {\bf 80}, 054917 (2009).

\bibitem{Won09b}
C. Y. Wong, Nonlin. Phenom. Complex Syst. {\bf 12}, 315  (2009),  [arXiv:0911.3583].

\bibitem{Won11}
C. Y. Wong, Phys. Rev. C {\bf 84}, 024901 (2011).

\bibitem{Won12}
 C. Y. Wong, Invited talk presented at the 35th Symposium on Nuclear Physics, Cocoyoc, Mexico, January 3, 2012, to be published in IOP Conference Series,
[arXiv:1203.4441 (2012)].

\bibitem{Shu07} E. Shuryak, Phys. Rec. C {\bf 76}, 047901 (2007).

\bibitem{Vol05}
S. A. Voloshin, Nucl. Phys.  A {\bf 749},  287  (2005).

\bibitem{Chi08}
C. B. Chiu and R. C. Hwa
Phys. Rev. C {\bf 79},  034901 (2009).

\bibitem{Hwa03}
R. C. Hwa and C. B. Yang, Phys.Rev. C {\bf 67} 034902  (2003); 
R. C. Hwa and Z. G. Tan, Phys. Rev. C {\bf 72},  057902 (2005);
R. C. Hwa and C. B. Yang, [nucl-th/0602024].

\bibitem{Chi05}
C. B. Chiu and R. C. Hwa
Phys. Rev. C {\bf 72},  034903 (2005).


\bibitem{Hwa07}
R. C. Hwa, Phys. Lett. {\bf B666}, 228 (2008).

\bibitem{Pan07} 
V. S. Pantuev, [arXiv:0710.1882].

\bibitem{Dum08}
A. Dumitru, F. Gelis, L. McLerran, and R. Venugopalan,
Nucl. Phys. A{\bf 810}, 91 (2008). 

\bibitem{Gav08}
S. Gavin, and G. Moschelli, J. Phys. G {\bf 35}, 104084 (2008).

\bibitem{Gav08a}
S. Gavin, L. McLerran, and G. Moschelli, Phys. Rev. C {\bf 79}, 051902 (2009).

\bibitem{Arm04} N. Armesto, C. A. Salgado, and U. A. Wiedemann,
Phys. Rev. Lett. {\bf 93}, 242301 (2004).

\bibitem{Rom07}
P. Romatschke, Phys. Rev. C {\bf 75} 014901  (2007).

\bibitem{Maj07} A. Majumder, B. Muller, and S. A. Bass,
  Phys. Rev. Lett. {\bf 99}, 042301 (2007).

\bibitem{Dum07} A. Dumitru, Y. Nara, B. Schenke, and M. Strickland,
  Phys. Rev. C {\bf 78}, 024909 (2008); B. Schenke, A. Dumitru,
  Y. Nara, and M. Strickland, J. Phys. G {\bf 35}, 104109 (2008).

\bibitem{Miz08}
R. Mizukawa, T. Hirano, M. Isse, Y. Nara, and A. Ohnishi,
J. Phys. G {\bf 35}, 104083 (2008).

\bibitem{Jia08} Jianyong Jia and R.. Lacey, 
 Phys. Rev. C {\bf 79}, 011901 (2009).

\bibitem{Jia08a} Jianyong Jia, 
Eur. Phys. J. C {\bf 61}, 255 (2009).

\bibitem{Ham10}
Y. Hama, R. P. G. Andrade, F. Grassi, W.-L. Qian, Talk presented at ISMD2010, 21-25 September, 2010, University of Antwerp (Belgium),  [arXiv:1012.1342]; R.P.G.Andrade, F.Gardim, F.Grassi, Y.Hama, W.L.Qian
 [arXiv:1107.0216].
 
\bibitem{Dum10}
A. Dumitru, K. Dusling, F. Gelis, J. Jalilian-Marian, T. Lappi, and R. Venugopalan,
Phys. Lett. {\bf B697}   21   (2011), [arXiv:1009.5295].

\bibitem{Wer11}
K. Werner, Iu. Karpenko, K. Mikhailov, and T. Pierog
[arXiv:1104.3269];
Fu-Ming Liu, K. Werner,
[arXiv:1106.5909].

\bibitem{Hwa11}
R. C. Hwa, C. B. Yang,
Phys. Rev.  C {\bf 83}, 024911 (2011), [arXiv:1011.0965]. 

\bibitem{Chi10}
C. B. Chiu and R. C. Hwa, [axriv:1012:3486].

\bibitem{Tra10}
T. A. Trainor, arXiv:1008.4757; T. A. Trainor, arXiv:1011.6351;
T. A. Trainor, arXiv:1012.2373.

\bibitem{Tra11a}
T. A. Trainor and D. T. Kettler, Phys. Rev.  C {\bf 83},
034903 (2011).

\bibitem{Tra11b}
T. A. Trainor and D. T. Kettler, arXiv:1010.3048.

\bibitem{Tra11c}
T. A. Trainor and R. L. Ray, 
arXiv:1105.5428; R. L. Ray, arXiv:1106.5023.

\bibitem{Sch11}
B. Schenke, [arXiv:1106.6012];
B. Schenke, S. Jeon, C. Gale, [arXiv:1109.6289].


\bibitem{Pet11}
H. Petersen, C. Greiner, V. Bhattacharya, S. A. Bass,
[arXiv:1105.0340].


\bibitem{Arb11}
B. A. Arbuzov, E. E. Boos, and V. I.  Savrin,
[arXiv:1104.1283].


\bibitem{Aza11}
M. Yu. Azarkin, I. M. Dremin, and A. V. Leonidov,
[arXiv:1102.3258].

\bibitem{Hov11}
H. R. Grigoryan, and Y. V. Kovchegov,
[arXiv:1012.5431].

\bibitem{Bau11}
I. Bautista, J. Dias de Deus, and C. Pajares,
[arXiv:1011.1870].

\bibitem{Che10}
 I. O. Cherednikov and N. G. Stefanis,
[arXiv:1010.4463].

 \bibitem{Dre10}
   I.  M. Dremin,  and V. T. Kim,
[arXiv:1010.0918].
 
\bibitem{Lev11} 
E. Levin, A. H. Rezaeian,
[arXiv:1105.3275]. 

\bibitem{Wan92}
X.  N.  Wang and M. Gyulassy, Phys. Rev.  D {\bf 44}, 3501 (1991);
X. N. Wang and M. Gyulassy, Phys. Rev. Lett. {\bf 68}, 1480 (1992). 


\bibitem{Bjo82}
J. D. Bjorken, Fermilab-Pub-82-059-THY (1982).


\bibitem{Gyu94}
M. Gyulassy and X. N. Wang, Nucl. Phys. {\bf B 420}, 583 (1994).


\bibitem{Bai96}
R. Baier, Yu. L. Dokshitzer, A. H. S. Peigne, and D. Schiff,
Nucl. Phys. {\bf B478}, 577 (1996).

\bibitem{Wie00}
U. A. Wiedemann, Nucl. Phys. {\bf B582}, 409 (2000).

\bibitem{Gyu01}
M. Gyulassy, P. Levai, and I. Vitev, Nucl. Phys. {\bf B594}, 371 (2001).

\bibitem{Djo04}
M. Djordjevic and M. Gyulassy,
Nucl. Phys.  {\bf A733},  265 (2004).



\bibitem{Adl04}
S. S. Adler $et~al.$ (PHENIX Collaboration), Phys. Rev. {\bf C69}, 034910 (2004); J. Adam $et~al.$ (STAR Collaboration),
Phys. Rev. Lett. {\bf 91}, 172302 (2003); B. B. Back $et~al.$ 
(PHOBOS  Collaboration), Phys. Lett. {\bf B578}, 297 (2004); 
I. Arsene $et~al.$ (BRAHMS Collaboration),
Phys. Rev. Lett. {\bf 91}, 072305 (2003); 
B. I. Abelev $et~al.$ (STAR Collaboration),
Phys. Rev. Lett. {\bf 98}, 192301 (2007);
S. S. Adare $et~al.$ (PHENIX Collaboration), Phys. Rev. Lett. {\bf 97}, 252002 (2006). 





\bibitem{Che69} H. Cheng and T. T. Wu, 
Phys. Rev.  {\bf 186}, 1611 (1969).

\bibitem{Che87} H. Cheng and T. T. Wu, {\it Expanding Protons: Scattering at High Energies} ,  M. I. T. Press, 1987.



\bibitem{Fen96}
Y.  J. Feng, O. Hamidi-Ravari, and C. S. Lam, Phys. Rev.  D {\bf 54}, 3114 (1996).

\bibitem{Fen97}
Y.  J. Feng, O. Hamidi-Ravari, and C. S. Lam, Phys. Rev.  D {\bf 55}, 4016 (1997).

\bibitem{Lam96a}
C. S. Lam,  Lectures given at the First Asia Pacific Workshop on Strong Interactions, Taipei. August 1st to 27th, 1996, hep-ph/9704240.

\bibitem{Lam97}
C. S. Lam and K.F. Liu, Nucl. Phys. {\bf B 483},   514 (1997).

\bibitem{Lam97a}
C. S. Lam and K.F. Liu, Phy. Rev. Lett. {\bf 79}, 597 (1997).

\bibitem{Lam97b}
C. S. Lam, Chin. Jour. Phys. {\bf 35},  758  (1997), hep-ph/9805210.


\bibitem{Ova11}
G. Ovanesyan and  I. Vitev	,
JHEP 1106 (2011) 080.

\bibitem{Meh11}	
Y. Mehtar-Tani, C.A. Salgado, and K. Tywoniuk,
 J.Phys.  {\bf G38},  124063  (2011).

\bibitem{Arm12}
N. Armesto, Hao Ma, Y. Mehtar-Tani, C. A. Salgado, K. Tywoniu,  JHEP {\bf 1201},  109  (2012),

\bibitem{Gla59}
R. J. Glauber, in {\it Lectures in Theoretical Physics}, edited by
W. E. Brittin and L. G. Dunham (Interscience, N.Y., 1959), Vol. 1,
p. 315.


\bibitem{Note}
In this manuscript, we use the Feynman rule in Ref.\ \cite{Che87} which includes explicitly an overall multiplicative phase factor of $(-i)$.  Such a overall phase $(-i)$ is not explicitly stated but  implicitly implied in Appendix A-4 of Ref.\ \cite{Itz80}. 

\bibitem{Itz80}
C. Itzykson and J.B. Zuber,  {\it Quantum Field Theory},  Mc-Graw Hill, (1980). 

\bibitem{Ber82}
V. B. Berestetskii, E. M. Lifshitz, and L. P. Pitaevskii, 
{\it Quantum Electrodynamics}, Pergamon Press, 1982.

\bibitem{Lo76}
C. Y. Lo and H. Cheng, Phys. Rev. D{\bf 13}, 1131 (1976);
P. S. Yeung, Phys. Rev. D 13, 2306–2317 (1976).

\bibitem{Che81}
H. Cheng,  J. A. Dickinson, and K. Olaussen, Phys. Rev. D {\bf 23}, 534 (1981).

\bibitem{CMS12b}
CMS Collboration, in CMS Physics Analaysis Summary HIN-11-006, available on the CERN CDS Information Server at 
http://cdsweb.cern.ch/record/1353583/files/HIN-11-006-pas.pdf.



\bibitem{Pes95}
M. E. Peskin and D. V. Schroeder
{\it An Introduction To Quantum Field Theory}, Addison-Wesley Publishing Company, (1995).

\bibitem{Gla70}
R. J. Glauber,
in {\it High-Energy Physics and Nuclear Structure}, edited by S. Devons, Press Press, N.Y. , 1970.
 
\bibitem{ATL11b}
G. Aad $et~ al.$ (Atlas Collaboration), Phys.  Rev.  Lett. {\bf 105}, 252303 (2010),
 arXiv:1011.6182.

\bibitem{CMS11b}
 S. Chatrchyan $et~al.$ (CMS Collaboration),  Phys.  Rev. {\bf C84} , 024906 (2011),  arXiv:1102.1957. 

\bibitem{Won79}
C. Y. Wong, Phys. Lett. B {\bf 88}, 39 (1979).

\bibitem{Oll92}
J. Y. Ollitrault, Phys. Rev. D {\bf 46}, 229 (1992). 

\bibitem{Vol96}
S. Voloshin, Y. Zhang, Z. Phys. Rev. C {\bf 70}, 665 (1996).

\end{thebibliography}
\end{document}